\documentclass[12pt, letter]{article}
\usepackage[english]{babel}
\usepackage[latin1]{inputenc}
\usepackage{amsmath}
\numberwithin{equation}{section}        % two level equations nr, where first number is the section
\numberwithin{table}{section}
\numberwithin{figure}{section}
\usepackage{amsthm}                     % provides an enhanced version of \newtheorem command for defning theorem-like environments.
\usepackage{amsfonts}                   % set of miscellaneous TeX fonts that augment the standard Computer Modern set normally distributed with TeX
\usepackage{apalike}

\usepackage{natbib}
\usepackage{mathtools}                  % an extension package to amsmath
\mathtoolsset{showonlyrefs}             % the cited refs are shown and the uncited are not shown
\usepackage{booktabs}
\usepackage{tikz}
\usepackage{pgfplots}
\usepackage{pgfplotstable}
\usepackage[colorlinks=true,linkcolor=blue,citecolor=blue]{hyperref}
\usepackage{lscape} %Produce landscape pages in a (mainly) portrait document.
\usepackage{color}
\usepackage{enumerate}                  % enumerate environment an optional argument which determines the style in which the counter is printed
\usepackage{setspace}                  % Deux lignes pour un espace interligne double
\usepackage{geometry}
\newtheorem{lemma}{Lemma}
\newtheorem{remark}{Remark}
\newtheorem{result}{Result}

\def\Ib{{\bf I}}
\def\Tb{{\bf T}}
\def\tb{{\bf t}}
\def\xb{{\bf x}}
\def\E{{\rm E}}
\def\V{{\rm V}}
\def\betab{\boldsymbol{\beta}} %$\beta$
\def\gammab{\boldsymbol{\gamma}} %$\gamma$
\def\lambdab{\boldsymbol{\lambda}} %$\lambda$

\title{Inference from Sampling with Response Probabilities Estimated via Calibration\footnote{Caren Hasler, Institut de Statistique, Universit\'e de Neuch\^atel, Av. de Bellevaux 51, 2000 Neuch\^atel, caren.hasler@unine.ch}}%
\author{}

\clearpage

\begin{document}

%\begin{bibunit}[apalike]

\maketitle

\setlength{\parindent}{0.5in}

\begin{abstract}
A solution to control for nonresponse bias consists of multiplying the design weights of respondents by the inverse of estimated response probabilities to compensate for the nonrespondents. Maximum likelihood and calibration are two approaches that can be applied to obtain estimated response probabilities. We consider a common framework in which these approaches can be compared. We develop an asymptotic study of the behavior of the resulting estimator when calibration is applied. A logistic regression model for the response probabilities is postulated. Missing at random and unclustered data are supposed. Three main contributions of this work are: 1) we show that the estimators with the response probabilities estimated via calibration are asymptotically equivalent to unbiased estimators and that a gain in efficiency is obtained when estimating the response probabilities via calibration as compared to the estimator with the true response probabilities, 2) we show that the estimators with the response probabilities estimated via calibration are doubly robust to model misspecification and explain why double robustness is not guaranteed when maximum likelihood is applied, and 3) we discuss and illustrate problems related to response probabilities estimation, namely existence of a solution to the estimating equations, problems of convergence, and extreme weights. We explain and illustrate why the first aforementioned problem is more likely with calibration than with maximum likelihood estimation. We present the results of a simulation study in order to illustrate these elements.
\end{abstract}


\clearpage

% ----------------------------------------------------------------
\section{Introduction}\label{section:introduction}
% ----------------------------------------------------------------

Under complete response the Horvitz-Thompson~\citeyearpar[HT,][]{hor:tho:52} estimator is unbiased. With nonresponse, however, this estimator is unavailable. Nonresponse can be seen as a second phase of the survey, where the mechanism that yields the nonresponse called the \emph{response mechanism} is unknown \citep{oh:sch:83,sar:swe:87}. If the response probabilities were known, a two-phase estimator with response probabilities as inclusion probabilities of the second phase would be unbiased. Unfortunately, the response probabilities are unknown in practice. A solution to control for nonresponse bias is to postulate a model for the response probabilities, estimate these probabilities based on the postulated model, and use the estimated response probabilities in a two-phase estimator. The resulting estimator is called \emph{two-phase Nonresponse Weighting Adjusted (NWA) estimator} or \emph{empirical double expansion estimator}. \cite{sar:lun:05} and \cite{haz:bea:17:weights:review} provide overviews of some NWA estimators and weighting systems adjusted for nonresponse.

Two general approaches to NWA estimators are Maximum Likelihood Estimation (MLE) and calibration \citep{dev:sar:92}. %and 3) the score method.
In the first approach, a model such as the logistic regression model is postulated~\citep{cas:sar:wre:83,ekh:laa:91}. The parameters of the model are estimated via MLE and fitted response probabilities are obtained based on the estimated parameters. In the second approach, calibration weights are found so that the resulting NWA estimator of some auxiliary variables is equal to its population total (calibration at the population level) or to its full sample HT estimator (calibration at the full sample level). The calibration weights can be viewed as the design weights times the inverse of the estimated response probabilities. To the best of our knowledge, the first author to suggest the use of what would later be called calibration weighting to estimate the response probabilities is \cite{fol:91}, shortly followed by \cite{dev:dup:93} and \cite{dup:93}. \cite{lun:sar:99} further study point and variance estimators for both levels of calibration, population and sample. %Some authors allow for the variables that explain the response probabilities to differ from the calibration variables. In this case, generalized calibration is used, such as for instance in \cite{dev:02}, \cite{kot:06}, \cite{cha:kot:08}, and \cite{les:haz:dha:19}. This goes beyond the scope of this work. %In the third approach, classes homogeneous with respect to the probability to respond are created and the response probabilities are estimated via the response rate of the classes \citep{lit:86,elt:yan:97,haz:bea:07}.

The first approach is studied in depth in \cite{kim:kim:07}, which presents asymptotic properties of the NWA estimator under a general response model. Two main results of their paper are: 1) the NWA estimator with response probabilities estimated via MLE is asymptotically equivalent to an unbiased estimator and 2) a gain in efficiency is obtained when estimating the response probabilities via MLE as compared to the estimator with the true response probabilities. The second result was also shown by \cite{bea:05a} under the logistic regression model.

The second approach can be divided into two levels: calibration at the sample level and calibration at the population level. The NWA estimator obtained when the response probabilities are estimated via calibration at the sample level is a particular case of the propensity-score-adjustment estimator of \cite{kim:rid:12}. These authors develop the asymptotic properties of this estimator in a theoretical framework different from that considered in \cite{kim:kim:07}. This estimator is also considered in \cite{ian:mil:fol:1991} which focuses on practical aspects of NWA estimation with calibration at the sample level. It does not provide any theory.

The main goal of both approaches is to reduce the nonresponse bias and, if possible, the variance of population estimators. The second approach, calibration, also ensures consistency between estimated and known population totals. This is not the case of the first approach, MLE. However, the second approach, i.e., direct estimation of the response probabilities via calibration, called \emph{one-step approach}, is sometimes criticized as it tends to yield biased estimates when the response model is misspecified \citep{haz:les:16}. An alternative consists of first estimating the response probability via MLE and then applying calibration to ensure consistency between estimated and known totals. This alternative is called \emph{two-step approach}. The reader may refer to \cite{haz:les:16} and \cite{haz:bea:17:weights:review}, p.222, for a discussion of the one- and two-step approaches.

In this paper, we study MLE and the one-step approach to calibration for nonresponse weighting adjustment. We build on \cite{kim:kim:07} and develop asymptotic properties of the NWA estimator under the second approach, calibration at both the sample and the population levels. For the first time, a common theoretical framework is considered for both approaches to NWA estimation, namely MLE and calibration. This allows us to compare the asymptotic behavior of the resulting NWA estimators in terms of bias and variance under common assumptions. We postulate a logistic regression model for the response probabilities. We suppose that the data are missing at random~\citep[see][for a detailed definition]{rub:76} and unclustered. Two main theoretical results are 1) the NWA estimators with the response probabilities estimated via calibration are asymptotically equivalent to unbiased estimators and 2) a gain in efficiency is obtained when estimating the response probabilities via calibration as compared to the estimator with the true response probabilities. These results are valid for both levels of calibration, population and full sample.

Another main contribution of this work is the study of the double-robustness of the NWA estimators. Indeed, both approaches assume, implicitly or explicitly, two models: 1) a model that links the variable of interest and the auxiliary variables, called \emph{superpopulation model}, and 2) a model for the response probabilities, called \emph{response model}. We show that the NWA estimators with response probabilities estimated via calibration are doubly robust. That is, these estimators are consistent even if one of the two aforementioned models is misspecified. We also explain why double robustness of the NWA estimator with response probabilities estimated via MLE is not guaranteed. To the best of our knowledge, only \cite{kot:lia:12} discusses double robustness of NWA estimation via calibration in probability sample surveys. In their article, the emphasis is put on an exponential form for the response probabilities. Finally, one last main contribution of this work is a discussion about problems of convergence and extreme weights. Indeed, it may happen that the estimating equations used to obtain estimated response probabilities do not admit a solution. This problem arises with calibration. In other cases, a solution to the estimating equations exists but the resulting weights, that is, the inverse of the estimated response probabilities, may be very large. We explain this phenomenon and give illustrative examples. Results of a simulation study confirm the theoretical results and practical considerations presented.

%We also show that the NWA estimator with response probabilities estimated via calibration at the population level is generally more efficient than the NWA estimator with response probabilities estimated via calibration at the sample level, but may be associated with problems. Indeed, the optimization program may fail to find a solution to the calibration equation or may return extreme weights for some combinations of sampling designs and response mechanisms. This is less likely to happen when calibrating at the sample level. We also show that if the variable of interest is a linear combination of the auxiliary variables, the NWA estimator with response probabilities estimated via calibration at the population level, respectively full sample level, is asymptotically equivalent to the true population total, respectively full sample HT estimator.
%
%\re{Discuss convergence problems, i.e. when a solution to the calibration equation does not exist.}

The paper is organized as follows: Section~\ref{section:framework} contains pieces of notation and important concepts. In Section~\ref{section:estimation}, we present both approaches to response probabilities estimation. We describe some asymptotic properties of the NWA estimators of interest in Section~\ref{section:asymptotics} with some technical elements left in the Appendix. We discuss double robustness to model misspecification in Section~\ref{section:asymptotics2}. Section~\ref{section:Convergence} contains a discussion about convergence to and existence of a solution to the extimating equations of response model parameters and extreme weights. In Sections~\ref{section:variance} and \ref{section:variance:estimation}, we present the variance and variance estimation of the NWA estimators of interest, respectively. Section~\ref{section:simulation} contains the results of a simulation study. A discussion closes the paper in Section~\ref{section:discussion}. The Appendix contains technical elements and the proofs of the stated results.

% ----------------------------------------------------------------
\section{Framework}\label{section:framework}
% ----------------------------------------------------------------

Consider a finite population $ U = \left\{1,2, \ldots , i, \ldots , N \right\}$ of size $N$. A vector of $v$ auxiliary variables $\xb_i = \left( x_{i1}, x_{i2}, \ldots, x_{iv}\right)$ is attached to a generic unit $i$. We suppose that the first auxiliary variable is constant and equal to 1. The parameter of interest is the population total
\begin{align}\label{equation:Y}
    Y = \sum_{i \in U} y_i,
\end{align}
for some unknown variable of interest $y$. A sample $s$ of size $n$ is selected from $U$ according to a non-informative probabilistic sampling design $p(\cdot)$ with the aim of observing $y_i$ for $i \in s$. A random sample $S$ is a random variable such that
$
  Pr(S = s) = p(s).
$
The random sample is also defined via an indicator variable $\left( a_i | i \in U\right)^\top$ where $a_i$ is 1 if unit $i$ is in the sample and 0 otherwise.
Consider
\begin{align}
    \pi_i = \Pr(i \in S) = \sum_{s \subset U;s \ni i}p\left( s \right),
\end{align}
the first-order inclusion probability of unit $i$ and suppose that $\pi_i > 0$ for all $ i \in U$. Let $\E_p(\cdot)$ and $\V_p(\cdot)$ denote the expectation and variance computed with respect to the sampling design $p(\cdot)$. Under complete response, the Horvitz-Thompson~\citeyearpar[HT,][]{hor:tho:52} estimator
\begin{align}\label{equation:ht}
    \widehat{Y}_\pi = \sum_{i \in S} \frac{y_i}{\pi_i}
\end{align}
is {design-unbiased} for $Y$, i.e., $\E_p(\widehat{Y}) = Y$.

Under nonresponse, each sampled unit $i \in S$ is classified as either \emph{respondent} or \emph{nonrespondent} depending on whether $y_i$ is observed or missing. Consider the response indicator vector $\left( r_i | i \in S\right)^\top$ where $r_i$ takes value 1 if $y_i$ is observed and 0 if it is missing and $p_i = \Pr(r_i = 1|i \in S)$ the response probability of a sampled  unit $i$. The set of respondents and nonrespondents are, respectively, $S_r = \left\{ i \in S | r_i = 1 \right\}$ of size $n_r$ and $S_m = \left\{ i \in S | r_i = 0 \right\}$. In the presence of nonresponse, the HT estimator in \eqref{equation:ht} is unavailable and the total $Y$ can be estimated via the \emph{two-phase} (or \emph{double expansion}) estimator
\begin{align}\label{eqn:ht:double}
    \widehat{Y}_{p} = \sum_{i \in S_r} \frac{y_i}{\pi_ip_i},
\end{align}
provided that $p_i > 0$ for all $i \in S$. This estimator is unbiased since $$\E_p\left\{E_q\left(\left.\widehat{Y}_{p}\right|S\right)\right\}=Y,$$ where $q(\cdot|S)$ is the probability distribution of $S_r$ given a sample $S$ and subscript $q$ indicates that the expectation is computed with respect to probability distribution $q(\cdot|S)$. The response probabilities are unknown in practice. To address this issue, a model for the response probabilities, called the \emph{response model}, is postulated. The response probabilities are estimated via this model, which yields estimated response probabilities $\widehat{p}_i$, and the \emph{NWA estimator} (or \emph{empirical double expansion estimator})
\begin{align}\label{eqn:ht:double:empirical}
            \widehat{Y}_{\widehat{p}} = \sum_{i \in S_r} \frac{y_i}{\pi_i\widehat{p}_i}
\end{align}
is used. The response probabilities are estimated via $\widehat{p}_i = f(\xb_i;\widehat{\lambdab})$ for some model $f(\xb_i;\lambdab)$ and estimator $\widehat{\lambdab}$ of $\lambdab$. A commonly used model for the response probabilities is the logistic regression model
\begin{align}\label{logistic}
  p_i = f(\xb_i;\lambdab) = \frac{\exp(\xb_i^\top \lambdab)}{1 + \exp(\xb_i^\top \lambdab)} = \frac{1}{1 + \exp(-\xb_i^\top \lambdab)},
\end{align}
where $\lambdab$ is a parameter vector to be estimated. Two available estimation methods are maximum likelihood and calibration, see Section~\ref{section:estimation}. Note that there are ways to use calibration weighting to adjust for nonresponse other than through an assumed logistic response model. For instance, other methods use a linear or logit function that bounds the probabilities of response between 0 and 1. More details can be found in \cite{dev:sar:92}, \cite{dev:sar:sau:93}, and \cite{haz:bea:17:weights:review}, among others. In the current work, we focus on the logistic regression model in \eqref{logistic}.

Some required assumptions on the response mechanism are:
\begin{enumerate}[({R}1):]
    \item \label{assumption:indep} The units respond independently of one another, i.e.
    \begin{align*}
        \Pr(i,j \in S_r|i,j \in S) = p_ip_j.
    \end{align*}
    \item \label{assumption:bounded:p} The response probabilities are bounded below, i.e. there exists a constant $c>0$ such that $p_i>c$ for all $i \in U$.
%    \item The propensity to respond independent from inclusion in sample, i.e.
%    \begin{align}
%          \Pr \left( i \in S_r| i \in S\right) = \Pr \left( i \in S_r \right)
%    \end{align}
    \item  \label{assumption:logistic} The response probabilities are $p_i = f(\xb_i, \lambdab^0)$ as defined in~\eqref{logistic} for some true unknown parameter vector $\lambdab^0$.
\end{enumerate}
%Assumptions (A\ref{assumption:MAR}) and (A\ref{assumption:indep}) are implicitly stated in Assumption~(A\ref{assumption:logistic}).
Assumption (R\ref{assumption:indep}) implies that each response indicators $r_i$ are draws of independent Bernoulli trials with parameters $p_i$, respectively. This means that $S_r$ is selected from $S$ via Poisson sampling design with inclusions probabilities $p_i$. Assumption (R\ref{assumption:logistic}) implies that
%\item \label{assumption:MAR}
the data are missing at random~\citep[see][for a detailed definition]{rub:76}. This means that
    \begin{align*}
        \Pr(i \in S_r|i \in S, \xb_i, y_i) = \Pr(i \in S_r|i \in S, \xb_i).
    \end{align*}
This means that the propensity to respond is independent from the variable of interest when the auxiliary variables are taken into account. This assumption may fail in practice when the propensity to respond still depends on the variable of interest when all available auxiliary information has been taken into account. If this is the case, one may use generalized calibration \citep{dev:02,kot:06,les:haz:dha:19,ran:mat:ner:23} to estimate the response probabilities instead of the approaches presented in Section~\ref{section:estimation}.

% ----------------------------------------------------------------
\section{Estimation}\label{section:estimation}
% ----------------------------------------------------------------

We consider two approaches to obtain the NWA estimator: MLE and calibration \citep{dev:sar:92}. \cite{kim:kim:07} study NWA estimators via MLE of the response probabilities under a general response model. For the logistic regression model, the maximum likelihood estimator of $\lambdab^0$ is the solution $\widehat{\lambdab}^{mle}$ to the estimating equation
\begin{align}\label{eqn:mle}
  Q^{mle} (\widehat{\lambdab}) = \sum_{i \in S} k_i\left\{r_i - f(\xb_i;\widehat{\lambdab})\right\}\xb_i = 0.
\end{align}
When $k_i=1$, the solution is the usual maximum likelihood estimator. When $k_i = 1/\pi_i$, we obtain a survey weighted estimating equation, which is often called \emph{pseudo-maximum likelihood}. The idea is that one first unbiasedly estimates the population likelihood estimating equation and then maximizes it. Other choices of $k_i$ are possible. We focus on the common two aforementioned choices. An efficiency gain of the NWA estimator in \eqref{eqn:ht:double:empirical} as compared to the two-phase estimator in \eqref{eqn:ht:double} with true response probabilities is claimed when $k_i = 1$ \citep{bea:05a,kim:kim:07}. This choice yields the best estimate of $\lambdab^0$ and of the response probabilities. The efficiency of the NWA estimator may, however, be improved upon with other choices of $k_i$, such as $k_i = 1/\pi_i$, for example. There is only very limited available literature on this choice. \cite{kot:12} discusses this choice and the impact on the efficiency of the NWA estimator for the case of response homogeneity groups. No general theory or guidelines about the choice of $k_i$ have been suggested yet in the literature. This goes beyond the scope of this paper.

Two levels of calibration are possible: calibration at the population level if the population total of the auxiliary variables is known and calibration at the sample level if the full sample HT estimator of the auxiliary variables is known. The calibration estimator of $\lambdab^0$ is the solution $\widehat{\lambdab}^{cal,U}$ to the estimating equation
\begin{align}\label{eqn:calU}
  \sum_{i \in S_r}  \frac{\xb_i}{\pi_i f(\xb_i;\widehat{\lambdab})}  = \sum_{i \in U}\xb_i,
\end{align}
or equivalently
\begin{align}\label{eqn:calU:S}
  Q^{cal,U}\left(\widehat{\lambdab}\right) = \sum_{i \in S_r}  \frac{\xb_i}{\pi_i f(\xb_i;\widehat{\lambdab})}  - \sum_{i \in U}\xb_i = 0,
\end{align}
if we calibrate at the population level. This calibration equation means that the response probabilities are chosen so that the NWA estimator of the auxiliary variables is equal to its population total. If we calibrate at the sample level, the calibration estimator of $\lambdab^0$ is the solution $\widehat{\lambdab}^{cal,S}$ to the estimating equation
\begin{align}\label{eqn:calS}
  \sum_{i \in S_r}  \frac{\xb_i}{\pi_i f(\xb_i;\widehat{\lambdab})}  = \sum_{i \in S}\frac{\xb_i}{\pi_i},
\end{align}
which is equivalent to
\begin{align}\label{eqn:calS:S}
  Q^{cal,S}\left(\widehat{\lambdab}\right) = \sum_{i \in S_r}  \frac{\xb_i}{\pi_i f(\xb_i;\widehat{\lambdab})}  - \sum_{i \in S}\frac{\xb_i }{\pi_i} = 0.
\end{align}
Estimating Equation~\eqref{eqn:calS} is suggested in \cite{ian:mil:fol:1991}. It means that the response probabilities are chosen so that the NWA estimator of the auxiliary variables is equal to its full sample HT estimator. Both estimating Equations~\eqref{eqn:calU} and \eqref{eqn:calS} can be solved using a software for calibration in the complete response case, such as function \emph{calib} of R package \emph{sampling} \citep{mat:til:21:RpackageSampling}.
%Indeed, noting that $1/f(\xb_i;\widehat{\lambdab}) = 1 + \exp(-\xb_i^\top \widehat{\lambdab})$, these equations can be written, respectively,
%\begin{align}
%  \sum_{i \in S_r}  \frac{\exp(-\xb_i^\top \widehat{\lambdab})}{\pi_i}\xb_i  &= \sum_{i \in U}\xb_i - \sum_{i \in S_r}\frac{\xb_i}{\pi_i},\\
%  \sum_{i \in S_r}  \frac{\exp(-\xb_i^\top \widehat{\lambdab})}{\pi_i}\xb_i  &= \sum_{i \in S\setminus S_r}\frac{\xb_i}{\pi_i}.
%\end{align}
%These are typical calibration equations with the raking ratio method, see~\cite{dev:sar:sau:93}.

When calibrating at the population level, the goal is to find weights, here response probabilities, so that the estimated total of some auxiliary variables matches the population total. When calibration at the sample level, the aim is to match the full sample HT estimator. Hence, the first approach attempts to correct for both the nonresponse and sampling error. The second approach attempts to only correct for the nonresponse error.

We compare four NWA estimators: 1) $\widehat{Y}_{\widehat{p}}^{mle,1}$ obtained with response probabilities estimated via Equation \eqref{eqn:mle} with $k_i =1$, 2) $\widehat{Y}_{\widehat{p}}^{mle,1/\pi}$ obtained with response probabilities estimated via Equation \eqref{eqn:mle} with $k_i = 1/\pi_i$, 3) $\widehat{Y}_{\widehat{p}}^{cal,U}$ obtained with response probabilities estimated via Equation \eqref{eqn:calU}, and 4) $\widehat{Y}_{\widehat{p}}^{cal,S}$ obtained with response probabilities estimated via Equation \eqref{eqn:calS}.

Both approaches, MLE and calibration, are here applied to estimate the response probabilities used in the NWA estimator in~\eqref{eqn:ht:double:empirical}. They differ, however, in spirit and required information in the estimation process. The spirit of MLE is to maximize the likelihood that the postulated response model generated the data at hand. The focus is the estimation of the response probabilities with no explicit parameter of interest in mind. Moreover, MLE does not explicitly assume a superpopulation model, i.e., a model that links the variable of interest and the auxiliary variables. We will see in Section \ref{section:asymptotics}, however, that MLE assumes an implicit superpopulation model. The idea of calibration is to find response probabilities so that the NWA estimators of the auxiliary variables match their population totals or full sample HT estimators. Hence, the spirit is to estimate the total of some auxiliary variables as precisely as possible so that the nonresponse bias of the total of the variable of interest is as small as possible when the variable of interest and the auxiliary variables are correlated. Calibration thus focuses on a particular parameter of interest, the total, and explicitly states a superpopulation model, a linear regression model.

Both approaches also differ in the required information in the estimation process. MLE requires to know the values $\xb_i$ for all sampled units $i \in S$. Calibration at the population level via estimating Equation~\eqref{eqn:calU} requires to know the values $\xb_i$ for all respondent units $i \in S_r$ and the population total of $\xb_i$. Calibration at the sample level via estimation Equation~\eqref{eqn:calS} requires to know the values $\xb_i$ for all respondent units $i \in S_r$ and the HT estimator of $\xb_i$ at the sample level. For MLE and calibration at the sample level, no information is needed about the $\xb_i$ out of the sample.

% ----------------------------------------------------------------
\section{Asymptotics I}\label{section:asymptotics}
% ----------------------------------------------------------------

\subsection{Theoretical Framework}

In this section, we build on the results and assumptions of \cite{kim:kim:07} to obtain some asymptotic properties of the NWA estimators obtained via calibration. We use the asymptotic framework of \cite{isa:ful:82}. Consider a sequence $U_N$ of embedded finite populations of size $N$ where $N$ grows to infinity. Consider a sequence of samples $s_N$ selected from $U_N$ with sampling design $p_N(\cdot)$. The first- and second-order inclusion probabilities associated with $p_N(\cdot)$ for some generic units $i$ and $j$ are $\pi_{N,i}$ and $\pi_{N,ij}$, respectively. In what follows, we will omit the subscript $N$ whenever possible to simplify notation. We consider the following common regularity conditions on the sequence of sampling designs to ensure consistent estimation of the HT estimator and its variance estimator.%, see for instance \cite{bre:ops:17:modelassist}.

\begin{enumerate}[({D}1):]
    \item\label{assumption:f} As $N \rightarrow + \infty$, we have $n/N \rightarrow \pi^* \in (0,1)$,
    \item\label{assumption:pii:bounded} For all $N$, $\pi_i > \lambda_1 > 0$ for all $i \in U$,
    \item\label{assumption:piij:bounded} For all $N$, $\pi_{ij} > \lambda_2 > 0$ for all $i,j \in U$,
    \item\label{assumption:deltaij} $\limsup\limits_{N\rightarrow + \infty} n \max\limits_{i,j\in U,i \neq j} \left|\pi_{ij} - \pi_i \pi_j \right| < + \infty$,
\end{enumerate}
where $\limsup$ is the limit superior. It is defined as the limit of the sequence of supremums. In the case of (D\ref{assumption:deltaij}), we can write
  \begin{align}
    \limsup\limits_{N\rightarrow + \infty} n \max\limits_{i,j\in U,i \neq j} \left|\pi_{ij} - \pi_i \pi_j \right|
            &= \lim\limits_{N\rightarrow + \infty} \sup\left\{\left. u_k     \right| k\geq N\right\},
  \end{align}
  where
  \begin{align}
    u_k &= n_k  \max\limits_{i,j\in U_k,i \neq j} \left|\pi_{k,ij} - \pi_{k,i} \pi_{k,j} \right|,
  \end{align}
  and $n_k$ is the size of $s_k$. Assumption (D\ref{assumption:deltaij}) states that the dependence between sample inclusion indicators is small enough \citep{bre:ops:17:modelassist}. Intuitively, if we regard $n \max\limits_{i,j\in U,i \neq j} \left|\pi_{ij} - \pi_i \pi_j \right|$ as a measure of dependence between the sample inclusion indicators, this measure should not increase to infinity. For instance, this assumption is satisfied for simple random sampling without replacement, Bernoulli sampling, and any stratified sampling that is not highly stratified. This assumption is not satisfied for cluster sampling or for highly stratified sampling designs. The next section summarizes the results of \cite{kim:kim:07} about the asymptotics of the NWA estimator when Maximum Likelihood is applied to obtain estimated response probabilities. The two sections that follow extend these results for the case in which calibration is used. In this section, the reference probability distribution for the convergence is the one jointly defined by the sampling mechanism and the response mechanism.

\subsection{Maximum Likelihood}

From Theorem 1 of \cite{kim:kim:07}, we have that under the regularity conditions (D\ref{assumption:f})-(D\ref{assumption:deltaij}), Assumptions~(R\ref{assumption:bounded:p})-(R\ref{assumption:logistic}) about the response mechanism, and additional regularity conditions (P\ref{assumption:bounded:moments})-(P\ref{assumption:lambda}) stated in the Appendix, the NWA estimator $\widehat{Y}_{\widehat{p}}^{mle}$ satisfies
    \begin{align}
  \frac{1}{N}\widehat{Y}^{mle}_{\widehat{p}} &= \frac{1}{N}\widehat{Y}^{mle}_{\widehat{p},l} + O_p(n^{-1}),
    \end{align}
where
    \begin{align}
  \widehat{Y}^{mle}_{\widehat{p},l} &=  \sum_{i \in S} \frac{1}{\pi_i} \left\{ k_i \pi_i p_i \xb_i^\top\gammab^{mle}_n + \frac{r_i}{p_i}
\left(y_i - k_i \pi_i p_i \xb_i^\top\gammab^{mle}_n\right)  \right\},\\
  \gammab^{mle}_n &= \left\{ \sum_{i \in S} k_i p_i(1-p_i)\xb_i\xb_i^\top \right\}^{-1}\sum_{i \in S} \frac{1-p_i}{\pi_i}\xb_iy_i.\\
    \end{align}
\begin{remark}\label{remark:nwa:mle:unbiased}
The NWA estimator $\widehat{Y}^{mle}_{\widehat{p}}$ behaves asymptotically like the linearized estimator $\widehat{Y}^{mle}_{\widehat{p},l}$, which is unbiased for the population total $Y$.
\end{remark}
\begin{remark}\label{remark:nwa:mle:linear}
If there exists a vector $\betab$ such that $y_i =  k_i \pi_i p_i \xb_i^\top \betab$ for all $i \in S$ then
    \begin{align}
    \widehat{Y}_{\widehat{p},l}^{mle} &= \sum_{i \in S} \frac{y_i}{\pi_i} .
    \end{align}
This means that $\widehat{Y}_{\widehat{p}}^{mle}$ is asymptotically equivalent to the full sample unknown HT estimator in this case. When estimating the response probability via MLE, see Equation~\eqref{eqn:mle}, we implicitly assume a superpopulation model, i.e., $y_i$ is a linear combination of $ k_i \pi_i p_i \xb_i$. %The resulting NWA estimator is such that the estimated response probabilities correct for the nonresponse error.
\end{remark}

\subsection{Calibration at the Sample Level}

\begin{result}\label{result:asymptotics:calS}
Let the sequence of sampling designs satisfy Assumptions~(D\ref{assumption:f})-(D\ref{assumption:deltaij}), the response mechanism satisfy Assumptions~(R\ref{assumption:bounded:p})-(R\ref{assumption:logistic}), and the sequence of finite populations satisfy Assumptions (P\ref{assumption:bounded:moments})-(P\ref{assumption:lambda}) in the Appendix. % on the response mechanism and the moments of the variables $y_i$ and $\xb_i$ hold.
The NWA estimator $\widehat{Y}_{\widehat{p}}^{cal,S}$ satisfies
    \begin{align}
  \frac{1}{N}\widehat{Y}_{\widehat{p}}^{cal,S} &= \frac{1}{N}\widehat{Y}_{\widehat{p},l}^{cal,S} + O_p(n^{-1}),
    \end{align}
where
    \begin{align}
  \widehat{Y}_{\widehat{p},l}^{cal,S} &= \sum_{i \in S} \frac{1}{\pi_i}\left\{  \xb_i^\top\gammab_{S} + \frac{r_i}{p_i}\left(y_i -  \xb_i^\top\gammab_{S}\right) \right\},\\
  \gammab_{S} &= \left( \sum_{i \in S} \frac{1-p_i}{\pi_i}\xb_i\xb_i^\top \right)^{-1}\sum_{i \in S} \frac{1-p_i}{\pi_i}\xb_iy_i.
    \end{align}
\end{result}
The proof is given in the Appendix.
\begin{remark}\label{remark:nwa:calS:unbiased}
The NWA estimator $\widehat{Y}^{cal,S}_{\widehat{p}}$ behaves asymptotically like the linearized estimator $\widehat{Y}^{cal,S}_{\widehat{p},l}$, which is unbiased for the population total $Y$.
\end{remark}
\begin{remark}\label{remark:nwa:calS}
If there exists a vector $\betab$ such that $y_i = \xb_i^\top \betab$ for all $i \in S$ then
    \begin{align}
    \widehat{Y}_{\widehat{p},l}^{cal,S} &= \sum_{i \in S} \frac{y_i}{\pi_i} .
    \end{align}
    This means that $\widehat{Y}_{\widehat{p}}^{cal,S}$ is asymptotically equivalent to the full sample unknown Horvitz-Thompson estimator in this case. When calibrating at the sample level via Equation~\eqref{eqn:calS}, we assume a superpopulation model, i.e., $y_i$ is a linear combination of $\xb_i$. %The resulting NWA estimator is such that the estimated response probabilities correct for the nonresponse error only.
\end{remark}

\subsection{Calibration at the Population Level}

\begin{result}\label{result:asymptotics:calU}
Let the sequence of sampling designs satisfy Assumptions~(D\ref{assumption:f})-(D\ref{assumption:deltaij}), the response mechanism satisfy Assumptions~(R\ref{assumption:bounded:p})-(R\ref{assumption:logistic}), and the sequence of finite populations satisfy Assumptions (P\ref{assumption:bounded:moments})-(P\ref{assumption:lambda}) in the Appendix. % on the response mechanism and the moments of the variables $y_i$ and $\xb_i$ hold.
The NWA estimator $\widehat{Y}_{\widehat{p}}^{cal,U}$ satisfies
    \begin{align}
  \frac{1}{N}\widehat{Y}_{\widehat{p}}^{cal,U} &= \frac{1}{N}\widehat{Y}_{\widehat{p},l}^{cal,U} + O_p(n^{-1}),
    \end{align}
where
    \begin{align}
  \widehat{Y}_{\widehat{p},l}^{cal,U} &=  \sum_{i \in U} \left\{  \xb_i^\top\gammab_{U} + \frac{a_i}{\pi_i}\frac{r_i}{p_i}\left(y_i -  \xb_i^\top\gammab_{U}\right)  \right\},\\
\gammab_{U} &= \left\{ \sum_{i \in U} (1-p_i)\xb_i\xb_i^\top \right\}^{-1} \sum_{i \in U} (1-p_i)\xb_i y_i.\\
    \end{align}
\end{result}
The proof is given in the Appendix.
\begin{remark}\label{remark:nwa:calU:unbiased}
The NWA estimator $\widehat{Y}^{cal,U}_{\widehat{p}}$ behaves asymptotically like the linearized estimator $\widehat{Y}^{cal,U}_{\widehat{p},l}$, which is unbiased for the population total $Y$.
\end{remark}
\begin{remark}\label{remark:nwa:calU:linear}
If there exists a vector $\betab$ such that $y_i = \xb_i^\top \betab$ for all $i \in U$ then
    \begin{align}
    \widehat{Y}_{\widehat{p},l}^{cal,U} &= \sum_{i \in U} y_i.
    \end{align}
This means that $\widehat{Y}_{\widehat{p}}^{cal,U}$ is asymptotically equivalent to the unknown population total in that case. When calibrating at the population level via Equation~\eqref{eqn:calU}, we assume a superpopulation model, i.e., $y_i$ is a linear combination of $\xb_i$. %The resulting NWA estimator is such that the estimated response probabilities correct for both the nonresponse and sampling error.
\end{remark}

\section{Asymptotics II: Double Robustness}\label{section:asymptotics2}
% ----------------------------------------------------------------

The results in Section~\ref{section:asymptotics} rely on Assumption~(R\ref{assumption:logistic}). That is, these results are valid if the response model is correctly satisfied. In this section, we show that the NWA estimators obtained with calibration may still be consistent when the response model is misspecified provided that a superpopulation model, i.e., a model that links the variable of interest to the auxiliary variables, is correctly specified. We say in this case that the resulting NWA estimators are doubly robust because consistency is maintained even when one of the two models, response model or superpopulation model, is misspecified. This is formalized by the results below. For the first result, two required assumptions about the response mechanism and estimated response probabilities are:
\begin{enumerate}[({R}1):]\setcounter{enumi}{3}
    \item \label{assumption:MAR} The data are MAR.
    \item \label{assumption:bounded:phat} The estimated response probabilities are bounded below, i.e., there exists a constant $c_1>0$ such that $\widehat{p}_i>c_1$ for all $i \in S$ and all $N$.
\end{enumerate}

\begin{result}\label{result:robustness:sup}
  Consider the superpopulation model $\xi: y_i = \xb_i^\top \betab + \varepsilon_i$ where $\E_{\xi}(\varepsilon_i) = 0$, $\E_{\xi}(\varepsilon_i\varepsilon_j) = \sigma^2 \leq +\infty$ if $i = j$ and 0 otherwise, and subscript $\xi$ means that the expectation and variance are computed with respect to model $\xi$. Suppose that assumptions (D\ref{assumption:f})-(D\ref{assumption:deltaij}), (R\ref{assumption:bounded:p}), (R\ref{assumption:MAR}), (R\ref{assumption:bounded:phat}) are satisfied. Then
  \begin{align}
    \frac{\widehat{Y}_{\widehat{p}}^{cal,U} - Y}{N} &= o_{\mathbb{P}}(1),\\
    \frac{\widehat{Y}_{\widehat{p}}^{cal,S} - Y}{N} &= o_{\mathbb{P}}(1).
  \end{align}
  Subscript $\mathbb{P}$ means that the reference probability distribution is that determined by the superpopulation model, the sampling design, and the response mechanism.
\end{result}
The proof is given in the Appendix. This result states that when the response probabilities are obtained via calibration, the resulting NWA estimators are consistent estimators of the true total. Result~\ref{result:robustness:sup} holds even when the response model in Assumption~(R\ref{assumption:logistic}) is misspecified.

\begin{result}\label{result:robustness:nr}
  Let the sequence of sampling designs satisfy Assumptions~(D\ref{assumption:f})-(D\ref{assumption:deltaij}), the response mechanism satisfy Assumptions~(R\ref{assumption:indep})-(R\ref{assumption:logistic}), and the sequence of finite populations satisfy Assumptions (P\ref{assumption:bounded:moments})-(P\ref{assumption:lambda}). Then
  \begin{align}
    \frac{\widehat{Y}_{\widehat{p}}^{cal,U} - Y}{N} &= o_{p}(1),\\
    \frac{\widehat{Y}_{\widehat{p}}^{cal,S} - Y}{N} &= o_{p}(1).
  \end{align}
\end{result}
The proof is given in the Appendix. This result states that when the response probabilities are obtained via calibration, the resulting NWA estimators are consistent estimators of the true total when the response model is correctly specified. Result~\ref{result:robustness:nr} holds even when the superpopulation model stated in Result~\ref{result:robustness:sup} is misspecified. Note that the probability distribution in Result~\ref{result:robustness:nr} is that determined by the sampling design and the response mechanism. The two quantities in Result~\ref{result:robustness:nr} are therefore also $o_{\mathbb{P}}(1)$.

From Results~\ref{result:robustness:sup} and \ref{result:robustness:nr}, we conclude that the NWA estimators obtained with calibration are doubly robust. That is, these estimators remain consistent even when one of the two models, superpopulation model or response model, is misspecified. When the response probabilities are estimated via MLE, however, consistency of the resulting NWA estimator is not guaranteed under the assumptions stated in the results. Indeed, when the response probabilities are obtained via MLE from Equation \eqref{eqn:mle}, the resulting weights may not be calibrated. This plays a central role in the proof of Results~\ref{result:robustness:sup}. As a result, if the double robustness of the NWA estimator obtained with MLE holds, further assumptions are required. This goes beyond the scope of this paper.

% ----------------------------------------------------------------
\section{Existence of a solution to the estimating equations, extreme weights, and convergence problems}\label{section:Convergence}
% ----------------------------------------------------------------

In some cases, the estimating equations used to obtain estimated response probabilities may not admit a solution. In other cases, a solution to the estimating equations exists but the resulting weights, that is, the inverse of the estimated response probabilities, may be very large. In this section, we explain this phenomenon and give illustrative examples. Note that in some cases, the optimization algorithm used to obtain the estimated response probabilities may not converge for numerical reasons.

A solution to the estimating equations may not exist when there are inconsistencies between the estimating equations of different auxiliary variables. This happens more frequently with calibration at the population level than with calibration at the sample level. It is not clear whether this may also happen when MLE is applied. We give examples of such inconsistencies in what follows. Simple random sampling is considered in order to simplify the explanation. With simple random sampling, the estimating equation for calibration at the population level can be written
\begin{align}\label{eqn:calU:srs}
  \frac{N}{n}\sum_{i \in S_r}  \frac{\xb_i}{ \widehat{p}_i}  = \sum_{i \in U}\xb_i.
\end{align}
Since the auxiliary variables contain a constant, the solution to this estimating equation must satisfy
\begin{align}
  \sum_{i \in S_r}  \frac{1}{ \widehat{p}_i}  = n.
\end{align}
Now suppose that the respondents' value for a given auxiliary variable are all larger than the population average value for this variable. That is, for an auxiliary variable $x$ there exists a constant $x_0$ such that $x_i > x_0$ for all $i \in S_r$ and $N^{-1}\sum_{i \in U} x_i < x_0$. For such an auxiliary variable, the estimating equation can be written
\begin{align}
  \frac{1}{n}\sum_{i \in S_r}  \frac{x_i}{ \widehat{p}_i}  = \frac{1}{N}\sum_{i \in U}x_i.
\end{align}
The left-hand side of this equation is strictly larger than $x_0$ since $x_i > x_0$ for all $i \in S_r$ and $\sum_{i \in S_r}  \frac{1}{ \widehat{p}_i}  = n$. The right-hand side is strictly smaller than $x_0$ since $N^{-1}\sum_{i \in U} x_i < x_0$. Therefore, it is not possible to find a solution that satisfies the estimating equations for both the constant auxiliary variable and auxiliary variable $x$. There is an inconsistency between the estimating equations of these variables. An example of such a case is if $x$ is a variable that takes value 1 if an individual is a male and 0 if the individual is a female, and if all respondents are males but that there are females in the population.

Such inconsistencies also happen with calibration at the sample level. Indeed, suppose that the respondents' value for a given auxiliary variable are all larger than the sample average value for this variable. That is, suppose that for an auxiliary variable $x$ there exists a constant $x_0$ such that $x_i > x_0$ for all $i \in S_r$ and $n^{-1}\sum_{i \in S} x_i < x_0$. As for calibration at the population level, it is in this case impossible to satisfy estimating equation~\eqref{eqn:calS} for both the constant auxiliary variable and auxiliary variable $x$. An example of such a case is if $x_i$ a variable that takes value 1 if an individual is a male and 0 if the individual is a female and if all respondents are males but there are females in the sample. It is not clear whether this may also happen when MLE is applied.

As we can see from the example given above, such inconsistencies are more likely to happen when calibrating at the population level than when calibrating at the sample level. Indeed, if all respondents are males but there are females in the sample, then there are also necessarily females in the population. This means that if there are inconsistencies with calibration at the sample level, then there are also inconsistencies with calibration at the population level. The opposite is not necessarily true. Indeed, if all respondents are males but there are females in the population, this does not necessarily mean that there are females in the sample.

In other cases, a solution to the estimating equations exists but the resulting weights, that is, the inverse of the estimated response probabilities, may be very large. This may for instance happen when there is an important imbalance in the respondents' values of the auxiliary as compared to the set on which we calibrate, i.e., the population or the sample. To illustrate this phenomenon, consider simple random sampling and calibration at the population level. Now suppose that the respondents' value for a given auxiliary variable are all larger than the population average value for this variable except for one respondents that has a value larger than the population average. This is not necessarily a case of inconsistency as presented above. However, in order to reweight the respondents so that Equation~\eqref{eqn:calU:srs} is satisfied, this particular respondents is reweighed so that it compensate all other respondents' values which are smaller than the population average. As a result, the associated weight $1/\widehat{p}_i$ for this particular respondent may be very large. This may also happen with calibration ar the sample level and with MLE. In Section~\ref{section:simulation:weight:convergence}, we discuss these problems of convergence and extreme weights through the results of a simulation study.

% ----------------------------------------------------------------
\section{Variance}\label{section:variance}
% ----------------------------------------------------------------

We suppose throughout this section that Assumption (R\ref{assumption:indep}) holds. Under nonresponse, we can write the variance of a generic estimator $\widehat{Y}_{g}$ as
\begin{align}
  \V\left(\widehat{Y}_{g}\right) &= \V_{sam}\left(\widehat{Y}_{g}\right) + \V_{nr}\left(\widehat{Y}_{g}\right),
\end{align}
where the two terms are the sampling variance and the nonresponse variance, respectively, and are given by
\begin{align}
  \V_{sam}\left(\widehat{Y}_{g}\right) &= \V_p\left\{ \E_q\left(\left. \widehat{Y}_{g} \right|S\right) \right\},\\
    \V_{nr}\left(\widehat{Y}_{g}\right)&= \E_p\left\{ \V_q\left(\left. \widehat{Y}_{g} \right|S\right) \right\}.
\end{align}

Based on this decomposition, the variance of the two-phase estimator $\widehat{Y}_{p}$ with the true response probabilities is given by
\begin{align}
    \V\left(\widehat{Y}_{p}\right) = \V_p\left( \sum_{i \in S} \frac{y_i}{\pi_i}  \right) + \E_p \left(  \sum_{i \in S} \frac{1}{\pi_i^2}\frac{1 - p_i}{p_i} y_i^2 \right).
\end{align}

Using the decomposition of the variance above, \cite{kim:kim:07}, p.507, write the variance of the linearized estimator $\widehat{Y}^{mle}_{\widehat{p},l}$ as
\begin{align}
  \V\left(\widehat{Y}^{mle}_{\widehat{p},l}\right) &= \V_{sam}\left(\widehat{Y}^{mle}_{\widehat{p},l}\right) + \V_{nr}\left(\widehat{Y}^{mle}_{\widehat{p},l}\right),
\end{align}
where
\begin{align}
  \V_{sam}\left(\widehat{Y}^{mle}_{\widehat{p},l}\right) &= \V_p\left( \sum_{i \in S} \frac{y_i}{\pi_i}  \right),\\
    \V_{nr}\left(\widehat{Y}^{mle}_{\widehat{p},l}\right) &= \E_p \left\{  \sum_{i \in S} \frac{1}{\pi_i^2}\frac{1 - p_i}{p_i} \left( y_i - k_i \pi_i p_i \xb_i^\top\gammab^{mle}_n \right)^2 \right\}.
\end{align}
The first term is the variance of the full sample HT estimator. The second term vanishes if there exists a vector $\betab$ such that $y_i = k_i \pi_i p_i \xb_i^\top \betab$. This agrees with Remark~\ref{remark:nwa:mle:linear} in Section~\ref{section:asymptotics} saying that $\widehat{Y}^{mle}_{\widehat{p},l}$ matches the full sample HT estimator when this relationship holds.

A similar decomposition holds for the case when calibration is applied. Indeed, we can write the variance of the linearized estimator $\widehat{Y}^{cal,S}_{\widehat{p},l}$ as
\begin{align}
  \V\left(\widehat{Y}^{cal,S}_{\widehat{p},l}\right) &= \V_{sam}\left(\widehat{Y}^{cal,S}_{\widehat{p},l}\right) + \V_{nr}\left(\widehat{Y}^{cal,S}_{\widehat{p},l}\right),
\end{align}
where
\begin{align}
  \V_{sam}\left(\widehat{Y}^{cal,S}_{\widehat{p},l}\right) &= \V_p\left( \sum_{i \in S} \frac{y_i}{\pi_i}  \right) ,\label{eqn:Vsam:calS}\\
    \V_{nr}\left(\widehat{Y}^{cal,S}_{\widehat{p},l}\right) &= \E_p \left\{  \sum_{i \in S} \frac{1}{\pi_i^2}\frac{1 - p_i}{p_i} \left(y_i -  \xb_i^\top\gammab_{S}\right)^2 \right\}.\label{eqn:Vnr:calS}
\end{align}
The first term is the variance of the full sample HT estimator. The second term vanishes if there exists a vector $\betab$ such that $y_i = \xb_i^\top\betab$. This agrees with Remark~\ref{remark:nwa:calS} saying that $\widehat{Y}^{cal,S}_{\widehat{p},l}$ matches the full sample HT estimator when this relationship holds.

Similarly, we can write
\begin{align}
  \V\left(\widehat{Y}^{cal,U}_{\widehat{p},l}\right) &= \V_{sam}\left(\widehat{Y}^{cal,U}_{\widehat{p},l}\right) + \V_{nr}\left(\widehat{Y}^{cal,U}_{\widehat{p},l}\right),
\end{align}
where
\begin{align}
  \V_{sam}\left(\widehat{Y}^{cal,U}_{\widehat{p},l}\right) &= \V_p\left\{ \sum_{i \in S} \frac{1}{\pi_i} \left(y_i -  \xb_i^\top\gammab_{U}\right) \right\},\label{eqn:Vsam:calU}\\
    \V_{nr}\left(\widehat{Y}^{cal,U}_{\widehat{p},l}\right) &= \E_p \left\{  \sum_{i \in S} \frac{1}{\pi_i^2}\frac{1 - p_i}{p_i} \left(y_i -  \xb_i^\top\gammab_{U}\right)^2 \right\}.\label{eqn:Vnr:calU}
\end{align}
The first term is the variance of the full sample HT estimator of the differences $y_i -  \xb_i^\top\gammab_{U}$. Both the first and second terms vanish if there exists a vector $\betab$ such that $y_i = \xb_i^\top\betab$. This agrees with Remark~\ref{remark:nwa:calU:linear} saying that $\widehat{Y}^{cal,U}_{\widehat{p},l}$ matches the true population total, which has zero variance, when this relationship holds.

 The decomposition of the variance of the estimators under study is summarized in Table~\ref{table:decomposition:variance}.
\begin{table}[tb]
\caption{Decomposition of the variance for four estimators. \label{table:decomposition:variance}}
\footnotesize
\begin{center}
\begin{tabular}{lcc}
\toprule
%        & \multicolumn{3}{c}{Comparison measure} \\
%        \cmidrule(r){2-4}
Estimator           & $\V_{sam}$ & $\V_{nr}$ \\
\midrule
$ \widehat{Y}_{p}$ & $\displaystyle \V_p\left( \sum_{i \in S} \frac{y_i}{\pi_i}  \right)$ &  $\displaystyle\E_p \left(  \sum_{i \in S} \frac{1}{\pi_i^2}\frac{1 - p_i}{p_i} y_i^2 \right)$\\
$\widehat{Y}^{mle}_{\widehat{p},l}$ & $\displaystyle\V_p\left( \sum_{i \in S} \frac{y_i}{\pi_i}  \right)$ & $ \displaystyle\E_p \left\{  \sum_{i \in S} \frac{1}{\pi_i^2}\frac{1 - p_i}{p_i} \left( y_i - k_i \pi_i p_i \xb_i^\top\gammab^{mle}_n \right)^2 \right\}$\\
$\widehat{Y}^{cal,S}_{\widehat{p},l}$ & $\displaystyle\V_p\left( \sum_{i \in S} \frac{y_i}{\pi_i}  \right)$ & $\displaystyle\E_p \left\{  \sum_{i \in S} \frac{1}{\pi_i^2}\frac{1 - p_i}{p_i} \left(y_i -  \xb_i^\top\gammab_{S}\right)^2 \right\}$\\
$\widehat{Y}^{cal,U}_{\widehat{p},l}$ & $\displaystyle\V_p\left\{ \sum_{i \in S} \frac{1}{\pi_i} \left(y_i -  \xb_i^\top\gammab_{U}\right) \right\}$  & $\displaystyle\E_p \left\{  \sum_{i \in S} \frac{1}{\pi_i^2}\frac{1 - p_i}{p_i} \left(y_i -  \xb_i^\top\gammab_{U}\right)^2 \right\}$\\
\bottomrule
\end{tabular}
\end{center}
\end{table}

\begin{remark}\label{remark:efficiency}
The sampling variance of the linearized estimators $\widehat{Y}^{mle}_{\widehat{p},l}$ and $\widehat{Y}^{cal,S}_{\widehat{p},l}$ is equal to the sampling variance of $\widehat{Y}_{p}$. Their nonresponse variance is no greater than that of $\widehat{Y}_{p}$. This means that the NWA estimators $\widehat{Y}^{mle}_{\widehat{p}}$ and $\widehat{Y}^{cal,S}_{\widehat{p}}$ are asymptotically equivalent to estimators that are at least as efficient as the estimator with the true response probabilities. This was shown in~\cite{kim:kim:07} for $\widehat{Y}^{mle}_{\widehat{p}}$, see p.505. In practice, this means that for large enough populations and samples we expect a gain in efficiency when estimating the response probabilities via MLE or calibration at the sample level as compared to using the true response probabilities.

We expect the sampling variance of the linearized estimator $\widehat{Y}^{cal,U}_{\widehat{p},l}$ to be smaller than the sampling variance of $\widehat{Y}_{p}$ provided that the residuals $y_i -  \xb_i^\top\gammab_{U}$ have less variation than the $y_i$'s. The nonresponse variance of $\widehat{Y}^{cal,U}_{\widehat{p},l}$ is no greater than that of $\widehat{Y}_{p}$. Thus, $\widehat{Y}^{cal,U}_{\widehat{p}}$ is asymptotically equivalent to an estimator that is at least as efficient as the estimator with the true response probabilities under the condition stated above. In practice, this means that for large enough populations and samples we expect a gain in efficiency when estimating the response probabilities via calibration at the population level as compared to using the true response probabilities provided that the residuals $y_i -  \xb_i^\top\gammab_{U}$ have less variation than the $y_i$'s.

Overall, there seems to be a gain in efficiency when using estimated response probabilities as compared to true response probabilities. A possible explanation is that estimating response probabilities can be viewed as a smoothing of the weights using an appropriate model. Such a smoothing has already been shown to improve the efficiency of the usual Horvitz-Thompson estimator, see \cite{bea:08:weights:biometrika} for instance.
\end{remark}

\begin{remark}\label{remark:efficiency:cal}
 Now comparing the variance of the NWA calibration estimators $\widehat{Y}^{cal,U}_{\widehat{p}}$ and $\widehat{Y}^{cal,S}_{\widehat{p}}$. We expect the sampling variance of the linearized estimator $\widehat{Y}^{cal,U}_{\widehat{p},l}$ to be smaller than the sampling variance of the linearized estimator $\widehat{Y}^{cal,S}_{\widehat{p},l}$ provided that the residuals $y_i -  \xb_i^\top\gammab_{U}$ have less variation than the $y_i$'s. Moreover, we expect the nonresponse variance of $\widehat{Y}^{cal,U}_{\widehat{p},l}$ to be close to that of $\widehat{Y}^{cal,S}_{\widehat{p},l}$, since the only difference is that the population coefficient $\gammab_{U}$ in the nonresponse variance of the former is replaced by a sample estimator $\gammab_{S}$ in the latter. In practice, this means that we expect a gain in efficiency of the NWA estimator when estimating the response probabilities via calibration at the population level as compared to the sample level, under the condition stated above about the residuals.
\end{remark}

% ----------------------------------------------------------------
\section{Variance Estimation}\label{section:variance:estimation}
% ----------------------------------------------------------------

We suppose throughout this section that Assumptions (D\ref{assumption:f})-(D\ref{assumption:deltaij}), (R\ref{assumption:indep})-(R\ref{assumption:logistic}), and (P\ref{assumption:bounded:moments})-(P\ref{assumption:lambda}) are satisfied. Using the decomposition of the variance, the following estimator may be used for the variance of the NWA estimator $\widehat{Y}^{mle}_{\widehat{p}}$, see \cite{kim:kim:07}, p.507,
\begin{align}
  \widehat{\V}\left(\widehat{Y}^{mle}_{\widehat{p}}\right) &= \widehat{\V}_{sam}\left(\widehat{Y}^{mle}_{\widehat{p},l}\right) + \widehat{\V}_{nr}\left(\widehat{Y}^{mle}_{\widehat{p},l}\right),
\end{align}
where
\begin{align}
  \widehat{\V}_{sam}\left(\widehat{Y}^{mle}_{\widehat{p},l}\right)    &= \sum_{i \in S_r}\frac{1-\pi_i}{\pi_i^2}\frac{y_i^2}{\widehat{p}_i} + \sum_{i,j \in S_r,i\neq j}\frac{\pi_{ij} - \pi_i\pi_j}{\pi_i\pi_j\pi_{ij}}\frac{y_i}{\widehat{p}_i}\frac{y_j}{\widehat{p}_j},\\
  \widehat{\V}_{nr} \left(\widehat{Y}^{mle}_{\widehat{p},l}\right)    &= \sum_{i \in S_r}\frac{1}{\pi_i^2}\frac{1-\widehat{p}_i}{\widehat{p}_i^2}\left(y_i - k_i \pi_i \widehat{p}_i \xb_i^\top \widehat{\gammab}^{mle}_n\right)^2,\\
 \widehat{\gammab}^{mle}_n &= \left\{ \sum_{i \in S_r} k_i(1-\widehat{p}_i)\xb_i\xb_i^\top \right\}^{-1}\sum_{i \in S_r} \frac{1}{\pi_i}\frac{1-\widehat{p}_i}{\widehat{p}_i}\xb_iy_i.\\
\end{align}

We consider the same approach to derive a variance estimator of NWA estimators $\widehat{Y}_{\widehat{p}}^{cal,U}$ and $\widehat{Y}_{\widehat{p}}^{cal,S}$.
Since $\widehat{Y}_{\widehat{p}}^{cal,U}$ is asymptotically equivalent to $\widehat{Y}_{\widehat{p},l}^{cal,U}$, we use
\begin{align}
  \widehat{\V}\left(\widehat{Y}_{\widehat{p}}^{cal,U}\right) &= \widehat{\V}_{sam}\left(\widehat{Y}_{\widehat{p},l}^{cal,U}\right) + \widehat{\V}_{nr}\left(\widehat{Y}_{\widehat{p},l}^{cal,U}\right),
\end{align}
where $\widehat{\V}_{sam}\left(\widehat{Y}_{\widehat{p},l}^{cal,U}\right)$ and $\widehat{\V}_{nr}\left(\widehat{Y}_{\widehat{p},l}^{cal,U}\right)$ are estimators of the variances in Equations~\eqref{eqn:Vsam:calU} and \eqref{eqn:Vnr:calU}, respectively. Under Assumptions~(D\ref{assumption:f})-(D\ref{assumption:deltaij}),(P\ref{assumption:bounded:moments}), estimator
\begin{align}
  \widehat{\V}\left(\widehat{Z}\right) &= \sum_{i,j \in S} \frac{\pi_{ij} - \pi_i\pi_j}{\pi_{ij}}\frac{z_i}{\pi_i}\frac{z_j}{\pi_j}
\end{align}
is design unbiased and consistent for the variance of a full sample HT estimator $\widehat{Z} = \sum_{i \in S}\pi_i^{-1}z_i$. Based on this formula, we can estimate the sampling variance via
\begin{align}
  \widehat{\V}_{sam}\left(\widehat{Y}_{\widehat{p},l}^{cal,U}\right) &=
                \sum_{i \in S_r} \frac{1-\pi_i}{\pi_i^2}\frac{e_i^2}{\widehat{p}_i}
                + \sum_{i,j \in S_r;i\neq j} \frac{\pi_{ij} - \pi_i\pi_j}{\pi_{ij}\pi_i\pi_j}\frac{e_i}{\widehat{p}_i}\frac{e_j}{\widehat{p}_j},\\
        e_i &= y_i -  \xb_i^\top\widehat{\gammab}^{cal}_n,\\
\widehat{\gammab}^{cal}_n &= \left( \sum_{i \in S_r} \frac{1}{\pi_i}\frac{1-\widehat{p}_i}{\widehat{p}_i}\xb_i\xb_i^\top \right)^{-1}\sum_{i \in S_r} \frac{1}{\pi_i}\frac{1-\widehat{p}_i}{\widehat{p}_i}\xb_iy_i,
\end{align}
where we substituted $\widehat{p}_i$ for the unknown $p_i$. Using the same substitution, we can estimate the second term via
\begin{align}
  \widehat{\V}_{nr}\left(\widehat{Y}^{cal,U}_{\widehat{p},l}\right) &=   \sum_{i \in S_r} \frac{1}{\pi_i^2}\frac{1 - \widehat{p}_i}{\widehat{p}_i^2} e_i^2 .
\end{align}

A similar construction for the variance of $\widehat{Y}_{\widehat{p}}^{cal,S}$ yields
\begin{align}
  \widehat{\V}\left(\widehat{Y}_{\widehat{p}}^{cal,S}\right) &= \widehat{\V}_{sam}\left(\widehat{Y}_{\widehat{p},l}^{cal,S}\right) + \widehat{\V}_{nr}\left(\widehat{Y}_{\widehat{p},l}^{cal,S}\right),
\end{align}
where $\widehat{\V}_{sam}\left(\widehat{Y}_{\widehat{p},l}^{cal,S}\right)$ and $\widehat{\V}_{nr}\left(\widehat{Y}_{\widehat{p},l}^{cal,S}\right)$ are estimators of the variances in Equations~\eqref{eqn:Vsam:calS} and \eqref{eqn:Vnr:calS}, respectively. We have
\begin{align}
  \widehat{\V}_{sam}\left(\widehat{Y}_{\widehat{p},l}^{cal,S}\right) &=
                \sum_{i \in S_r} \frac{1-\pi_i}{\pi_i^2}\frac{y_i^2}{\widehat{p}_i}
                + \sum_{i,j \in S_r;i\neq j} \frac{\pi_{ij} - \pi_i\pi_j}{\pi_{ij}\pi_i\pi_j}\frac{y_i}{\widehat{p}_i}\frac{y_j}{\widehat{p}_j},\\
  \widehat{\V}_{nr}\left(\widehat{Y}^{cal,S}_{\widehat{p},l}\right) &=   \sum_{i \in S_r} \frac{1}{\pi_i^2}\frac{1 - \widehat{p}_i}{\widehat{p}_i^2} \left( y_i -  \xb_i^\top\widehat{\gammab}^{cal}_n\right)^2.
\end{align}

% ----------------------------------------------------------------
\section{Simulation Study}\label{section:simulation}
% ----------------------------------------------------------------

\subsection{Simulation Settings}

Five different populations are considered and obtained as follows. For each population, we generate $N=2000$ population units. The auxiliary variables are the same across all five populations and are $\xb_i = \left(1, x_{i}\right)^\top$ where $x_{i}$ are observations of independent and identically distributed (iid) uniform random variables with parameters, i.e., bounds, 0 and 100. The values of the variables of interest are obtained as follows:
\begin{align}
  y_{1i} &= 1000 + 20x + \varepsilon_{1i},\\
  y_{2i} &= 1500 + 500\exp(-10 + 0.1x) + \varepsilon_{2i},\\
  y_{3i} &= \left\{
             \begin{array}{ll}
               1 & \hbox{with probability $\phi_i$,} \\
               0 & \hbox{otherwise,}
             \end{array}
           \right.
        \quad\mbox{where}\quad \phi_i =
        \left\{
          \begin{array}{ll}
            0.8 & \hbox{if $x_i > 75$,} \\
            0.2 & \hbox{otherwise,}
          \end{array}
        \right.\\
    y_{4i} &= 1000 + \varepsilon_{4i},\\
    y_{5i} &= 1000 + 20x + \varepsilon_{5i},
\end{align}
where $\varepsilon_{1i}, \varepsilon_{2i}, \varepsilon_{4i},$ and $\varepsilon_{5i}$ are observations if iid random normal distributions with mean 0 and standard deviation 750, 100, 750, and 50, respectively. In population  1, there is a linear relationship between $x$ and $y_1$ with a correlation of approximately $0.6$. In population 2, there is a non-linear relationship between $x$ and $y_2$. In population 3, $y_3$ is categorical and the values are obtained from independent Bernoulli random variables with parameter 0.8 for large values of $x$ and 0.2 for small values of $x$. In population 4, there is no relationship between $x$ and $y_4$. In population 5, there is a very strong linear relationship between $x$ and $y_5$ with a correlation of approximately 0.99.

\begin{figure}[!htb]
\centering
% The output from tikz()
% is imported here.
\input{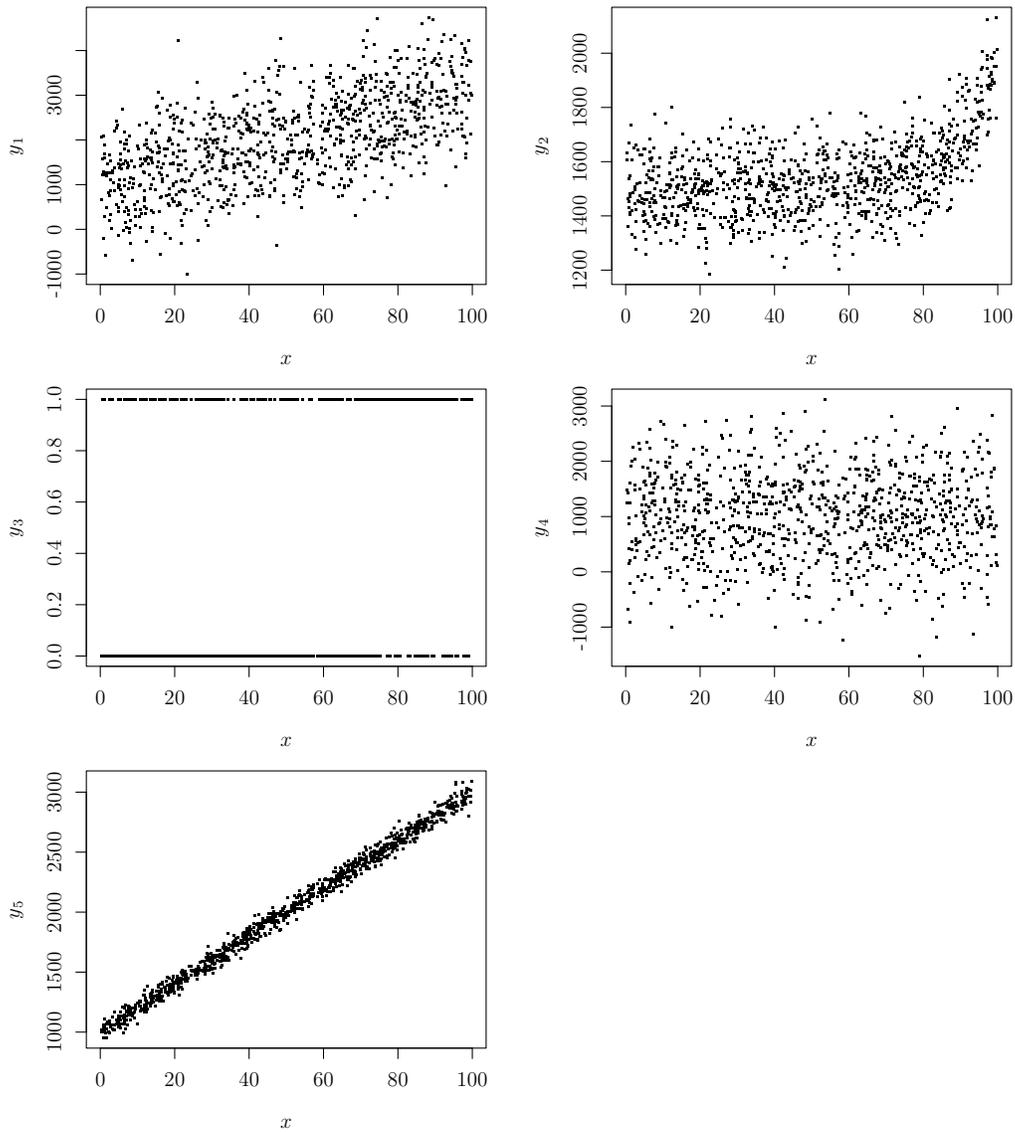}
\caption{Five populations.\label{fig:sim:pop}}
\end{figure}

\begin{figure}[!htb]
\centering
% The output from tikz()
% is imported here.
\input{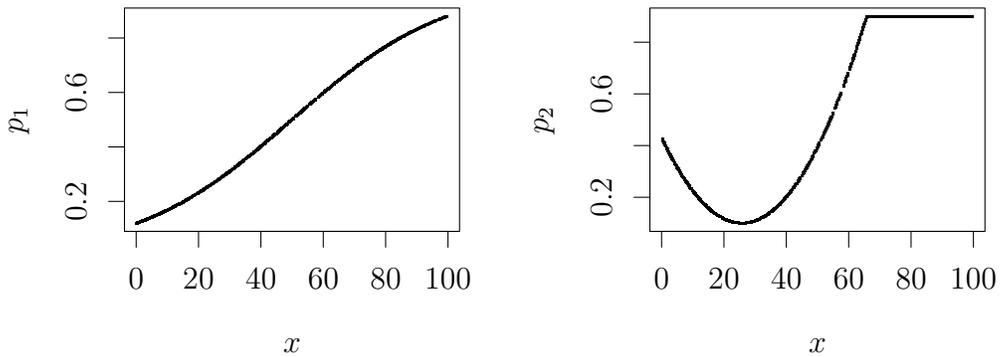}
\caption{Two vectors of response probabilities.\label{fig:sim:prob}}
\end{figure}

Two vectors of response probabilities are created as follows
\begin{align}
    p_{1i} &= \frac{1}{1 + \exp(-\xb_i^\top \lambdab)},\\
    p_{2i} &= \left\{
               \begin{array}{ll}
                 1 - a_1(x_i - k_1)^2 + h_1 & \hbox{if $a_1(x_i - k_1)^2 + h_1 > 0.01$,} \\
                 0.9 & \hbox{otherwise,}
               \end{array}
             \right.
\end{align}
where $a_1 = -0.0005$, $k_1 = 25.79116$, $h_1 = 0.9$, and $\lambdab = (-2,0.04)^\top$. Both vectors are constructed so that they yield a population mean response rate of approximately 50\%. Note that depending on the selected sample, the sample mean response rate may be larger or smaller than 50\% as units are not necessarily selected uniformly across all values of $x$. For the first vector of response probabilities, the logistic regression model in Equation~\eqref{logistic} is correctly specified. For the second vector of response probabilities, this model is misspecified. For both vectors, large values of $x$ tend to have large response probabilities. Figure~\ref{fig:sim:pop} shows the five populations and Figure~\ref{fig:sim:prob} the response probabilities as a function of the values of $x$.

% Open R and make the graphs than include in the paper. Next include and comment the results of the simulations.

Two sampling designs are considered: 1) simple random sampling with replacement where $n=200$ units are selected; 2) stratified sampling where two strata are considered. The first stratum contains the units with a $x$-value smaller than the median value of $x$, the second stratum contains those units with a $x$-value larger the median. Forty units are selected from the first stratum using simple random sampling. The sampling fraction in the first stratum is 4\%. One hundred and sixty units are selected from the second stratum using simple random sampling. The sampling fraction in the second stratum is 16\%.

Ten thousand simulations are run as explained in what follows for each population, each sampling design, and each vector of response probabilities. This results in 20 scenarios. A sample of size $n = 200$ is selected according to the sampling design. A set of respondents is generated with Poisson sampling design with the vector of response probabilities. Function \texttt{optim} is used to solve the estimating equations to obtain the parameters of the response model as presented in Section~\ref{section:estimation}. The function minimizes the maximum of the absolute relative value of the left-hand-side of estimating equations \eqref{eqn:mle}, \eqref{eqn:calU:S}, and \eqref{eqn:calS:S} over the auxiliary variables. We define that the algorithm converges if this maximum is less than 0.01. The initial value of the parameter vector is set to $(0,0)$ so that the initial response probabilities are all 1/2. When comparing the performance of the NWA estimators and their variance estimators, only those simulation runs for which the algorithm converges are used for computing comparison measures of a given estimator. The total $Y$ is estimated via seven estimators listed below.

\begin{enumerate}
  \item $\widehat{Y}$ (HT): the Horvitz-Thompson estimator. Note that this estimator is unavailable in practice with nonresponse. It serves here as a comparison point.
    \item $\widehat{Y}_{p}$ ($p$): estimator with the true response probabilities in~\eqref{eqn:ht:double}. This estimator is unavailable in practice. It serves here as a comparison point.
    \item $\widehat{Y}_{naive}$ (naive): estimator that ignores nonresponse, that is $\displaystyle\widehat{Y}_{naive} = \frac{n}{n_r} \sum_{i \in S_r} \frac{y_i}{\pi_i}$.
    \item $\widehat{Y}_{\widehat{p}}^{mle,1}$ (mle, $1$): NWA estimator with response probabilities estimated via MLE, Equation \eqref{eqn:mle}, with $k_i =1$.
    \item $\widehat{Y}_{\widehat{p}}^{mle,1/\pi}$ (mle, $1/\pi$): NWA estimator with response probabilities estimated via MLE, Equation \eqref{eqn:mle}, with $k_i = 1/\pi_i$.
    \item $\widehat{Y}_{\widehat{p}}^{cal,U}$ (cal, $U$): NWA estimator with response probabilities estimated via calibration at the population level, Equation \eqref{eqn:calU}.
    \item $\widehat{Y}_{\widehat{p}}^{cal,S}$ (cal, $S$): NWA estimator with response probabilities estimated via calibration at the sample level, Equation \eqref{eqn:calS}.
\end{enumerate}

\subsection{Performance of the NWA Estimators}

The performance of the estimators is assessed through the following comparison measures defined for a generic estimator $\widehat{Y}_g$:
\begin{itemize}
    \item   Absolute Monte Carlo relative bias ($|\mbox{RB}|$) defined as
            \begin{align}
                |\mbox{RB}| = \left|\frac{B}{Y}\right|,
            \end{align}
            where $\mbox{B} = \widehat{Y}_g^{(\cdot)} - Y$, $\widehat{Y}_g^{(\cdot)}$ is the mean of the estimator over the $L$ simulation runs (or the $L$ simulation runs for which the optimization algorithm converges if $\widehat{Y}_g$ is a NWA estimator),
            \begin{align}
                \widehat{Y}_g^{(\cdot)} = \frac{1}{L}\sum_{\ell = 1}^L \widehat{Y}_g^{(\ell)},
            \end{align}
            and $\widehat{Y}_g^{(\ell)}$ is the estimator $\widehat{Y}_g$ obtained at the $\ell$-th simulation,
    \item   Monte Carlo relative standard deviation (RSd) defined as
            \begin{align}
                \mbox{RSd} = \frac{\left( \mbox{VAR}\right)^{1/2}}{Y},
            \end{align}
            where
            \begin{align}
                \mbox{VAR} =  \frac{1}{L - 1} \sum_{\ell = 1}^L \left( \widehat{Y}_g^{(\ell)} - \widehat{Y}_g^{(\cdot)} \right)^2 .
            \end{align}
%    \item   the Monte Carlo relative root mean square error defined as
%            \begin{align}
%                \mbox{RRMSE} = \frac{\left(\mbox{B}^2 + \mbox{VAR} \right)^{1/2}}{Y}.
%            \end{align}
\end{itemize}

\begin{figure}[!htb]
\centering
% The output from tikz()
% is imported here.
\input{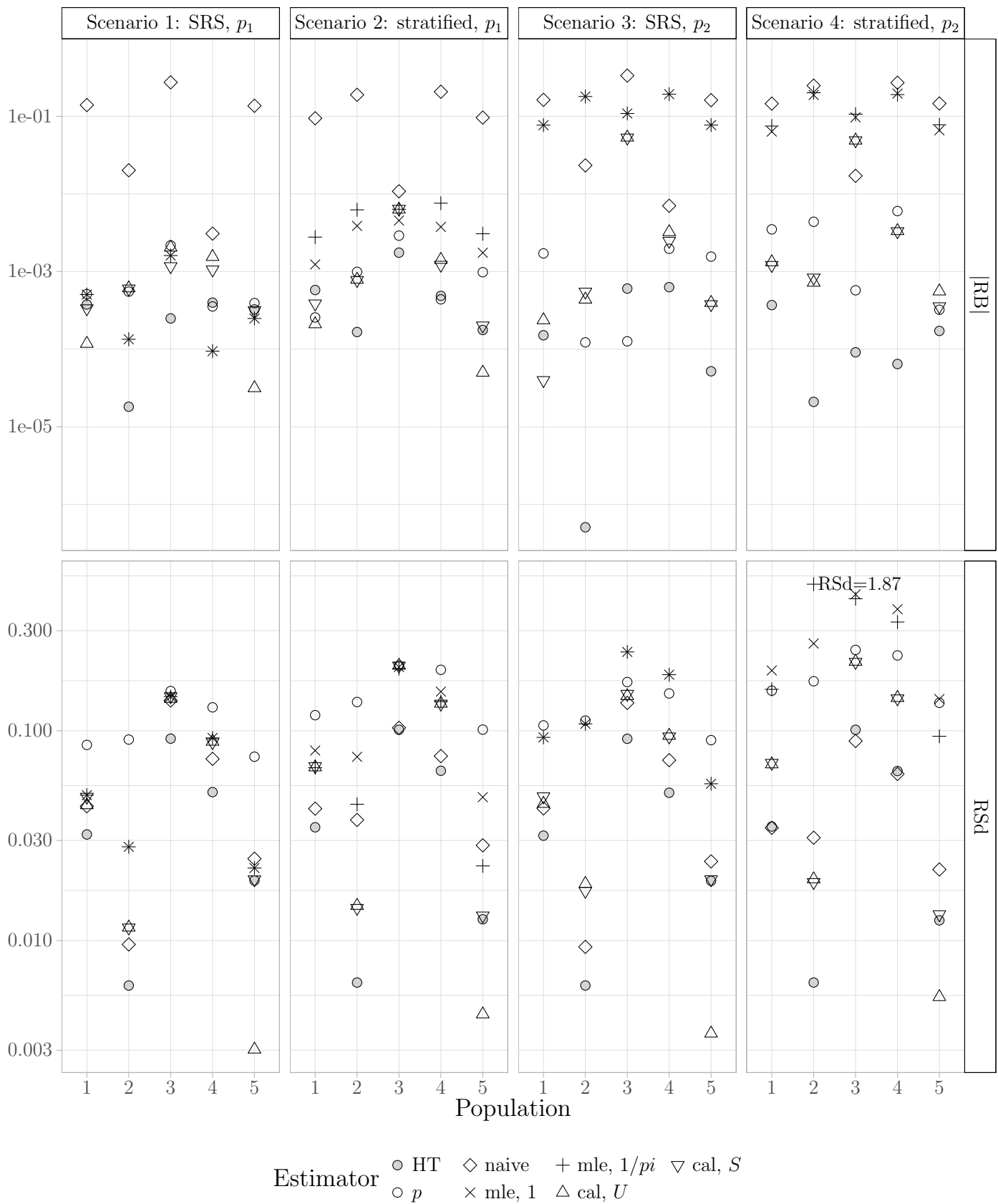}
\caption{$|\mbox{RB}|$ and RSd for seven estimators, five populations, and 4 scenarios.\label{fig:total}}
\end{figure}

The results are presented in Figure~\ref{fig:total}. The y-axes are displayed in logarithmic scales. For the plots of RSd, the maximum value on the $y$-axis is set to 0.5 for clarity reasons. One estimator has a value larger than 0.5 in scenario 4, population 2. This value is labelled on the graph. In scenarios 1 and 2, when the model for the response probabilities is correctly specified, all four NWA estimators show a RB of the same order as the RB of the HT estimator and the estimator with the true response probabilities $\widehat{Y}_p$. These last two estimators being unbiased, this result illustrates how the four NWA estimators are nearly unbiased, see Remarks~\ref{remark:nwa:mle:unbiased}, \ref{remark:nwa:calS:unbiased}, and \ref{remark:nwa:calU:unbiased}. In scenarios 3 and 4, when the model for the response probabilities is incorrectly specified, the two NWA estimators with response probabilities estimated via calibration show a RB of the same order as the RB of the HT estimator and the estimator with the true response probabilities $\widehat{Y}_p$. The two estimators with response probabilities estimated via MLE show a larger RB. This illustrates how calibration may provide a stronger protection against misspecification of the model for the response probabilities as compared to MLE. In all four scenarios, the naive estimator yields the larger RB.

\newgeometry{left=0.6in,right=0.7in, top =1in, bottom =1in}
\thispagestyle{empty}
%\begin{landscape}
%\input{figs/table_tot_srs_pois_clean.tex}
%\input{figs/table_tot_srs.tex}
%\input{figs/table_tot_pois.tex}
%\end{landscape}
\restoregeometry

In scenarios 1 and 2, when the model for the response probabilities is correctly specified, all four NWA estimators show a variance smaller than the variance of the estimator with the true response probabilities $\widehat{Y}_p$. This confirms that a gain in efficiency of the total estimator is obtained when estimating the response probabilities via MLE or calibration as compared to using the true response probabilities, see Remark~\ref{remark:efficiency}. In these two scenarios, all four NWA estimators show a RSd of the same order. In scenarios 3 and 4, when the model for the response probabilities is incorrectly specified, the two NWA estimators with response probabilities estimated via calibration show a RSd smaller than the RSd of the two NWA estimators with response probabilities estimated via MLE. This illustrates how calibration may provide a stronger protection against misspecification of the model for the response probabilities as compared to MLE.

%Third, estimator $\widehat{Y}_{\widehat{p}}^{cal,U}$ shows the best results in terms of variance overall. The stronger the strength of the linear relationship between the variable of interest and the auxiliary variables the smaller the variance of $\widehat{Y}_{\widehat{p}}^{cal,U}$. This is explained by the fact that when the linear relationship is strong, the variance of the residuals $y_i -  \xb_i^\top\gammab_{U} $ is small and therefore $\widehat{Y}_{\widehat{p}}^{cal,U}$ has a smaller variance than the other considered estimators, see Remark~\ref{remark:efficiency}. Last, a choice between $k_i = 1$ and $k_i = 1/\pi_i$ in MLE is not clear based on these results. For simple random sampling, both options yield equal performance. This was expected since all $1/\pi_i$ are equal with this sampling design. For Poisson sampling design, the estimator with $k_i = 1/\pi_i$ has lower variance but higher bias than the estimator with $k_i = 1$.

\subsection{Performance of the Variance Estimators}

The variance of the four NWA estimators is estimated for each simulation run with the formulae of Section \ref{section:variance:estimation}. The performance of the variance estimators is assessed through the following comparison measures defined for a generic estimator $\widehat{Y}_g$:
\begin{itemize}
    \item  Absolute Monte Carlo relative bias ($|\mbox{RB}|$) defined as
            \begin{align}
                |\mbox{RB}| = \frac{|B|}{\V_{sim}\left(\widehat{Y}_g\right)},
            \end{align}
            where $\V_{sim}\left(\widehat{Y}_g\right)$ is the variance of $\widehat{Y}_g$ over the $L$ simulation runs for which the optimization algorithm converges, $\mbox{B} = \widehat{\V}\left(\widehat{Y}_g^{(\cdot)}\right) - \V_{sim}\left(\widehat{Y}_g\right)$, and $\widehat{\V}\left(\widehat{Y}_g^{(\cdot)}\right)$ is the mean of $\widehat{\V}\left(\widehat{Y}_g\right)$ over the $L$ simulation runs,
%    \item   Mean length of the 95\% confidence interval (CI) based on the normal distribution
%            \begin{align}
%                95\% \mbox{CI} = \left[ \widehat{Y}_g - 1.96 \sqrt{\widehat{\V}\left(\widehat{Y}_g\right)} ; \widehat{Y}_g + 1.96 \sqrt{\widehat{\V}\left(\widehat{Y}_g\right)} \right]
%            \end{align}
%            over the $L = 10,000$ simulations,
        \item CR: the actual coverage rate of the 95\% confidence interval, i.e., the proportion of simulation runs for which the 95\% confidence interval contains the true total $Y$.
\end{itemize}

\begin{figure}[!htb]
\centering
% The output from tikz()
% is imported here.
\input{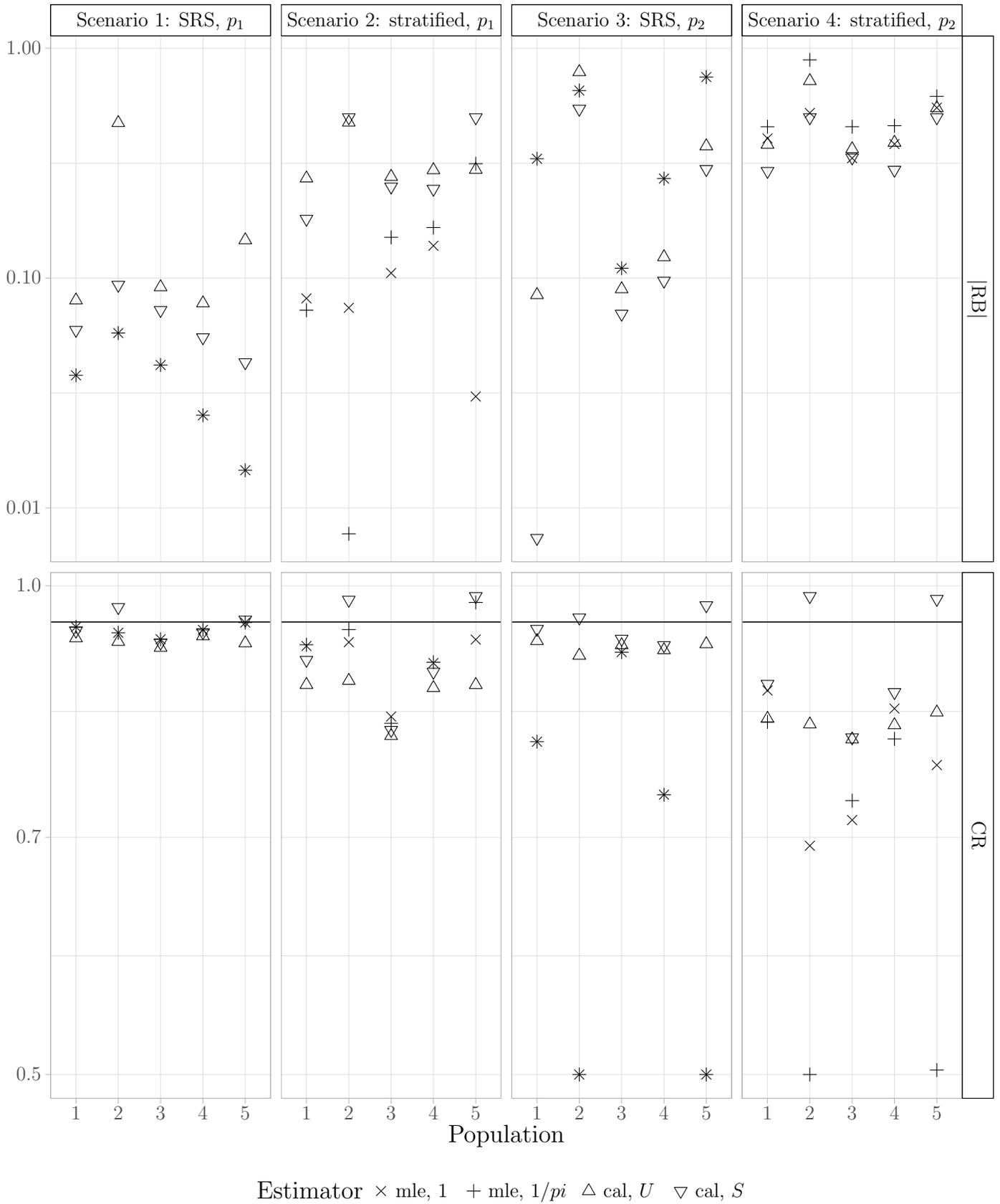}
\caption{$|\mbox{RB}|$ and CR for four variance estimators, five populations, and 4 scenarios.\label{fig:variance}}
\end{figure}

The results are presented in Figure~\ref{fig:variance}. The y-axes are displayed in logarithmic scales. To ease the reading of the graphs, four RB larger than 1 were set to 1 and five CR smaller than 0.5 were set to 0.5. In scenarios 1 and 2, when the model for the response probabilities is correctly specified, the RB of the variance estimator with response probabilities estimated via MLE tends to be smaller than the RB of the variance estimator with response probabilities estimated via calibration. In scenarios 3 and 4, when the model for the response probabilities is incorrectly specified, it is the opposite. In scenarios 1 and 2, all four variance estimators yield a CR generally close to the nominal coverage of 95\%. In scenarios 3 and 4, the variance estimator with response probabilities estimated via MLE yields very low CR in several cases.
%
%For simple random sampling, all four estimators yield comparable results in Population 1, with low RB and CR close or equal to the nominal coverage rate of 95\%. For the other two populations, when the correlation is weaker, all NWA estimators underestimate the variance (RB is negative) and yield a CR between 88\% and 93\%. For Poisson sampling, the estimator of the variance of $\widehat{Y}_{\widehat{p}}^{cal,U}$ underestimates the variance, while the other three estimators overestimate it. The length of the CI is much less with this estimator than with the other three and the CR tends to be slightly less. This means that estimator $\widehat{Y}_{\widehat{p}}^{cal,U}$ is preferred if we want a narrower CI with a CR slightly lower than the nominal coverage of 95\%, and the other three are preferred if we want a CR close to the nominal coverage of 95\%. The performance of all four estimators is therefore comparable. Overall, all variance estimators perform quite well for all three populations and both sampling designs.
%\input{figs/table_var_clean.tex}
\newgeometry{left=0.6in,right=0.7in, top =1in, bottom =1in}
\thispagestyle{empty}
%\begin{landscape}
%\input{figs/table_tot_srs_pois_clean.tex}
%\input{figs/table_tot_srs.tex}
%\input{figs/table_tot_pois.tex}
%\end{landscape}
\restoregeometry

\subsection{Weights and Convergence}\label{section:simulation:weight:convergence}

In order to illustrate the problems of convergence and extreme weights raised in Section~\ref{section:Convergence}, the following three comparisons measures are computed for each NWA estimator
\begin{itemize}
    \item   Maximum weight: the largest final adjusted weight $1/(\pi_i \widehat{p}_i)$ over all 10,000 simulations,
    \item   Mean Relative Error (Mean RE): the mean over 10,000 simulations of the maximum of the absolute relative error of the estimating equation,
    \item   Rate calib: the proportion of simulations for which the Mean RE is smaller than the threshold 0.01. We define that the algorithm converges to a solution when the Mean RE is smaller than this threshold.
\end{itemize}

\begin{figure}[!htb]
\centering
% The output from tikz()
% is imported here.
\input{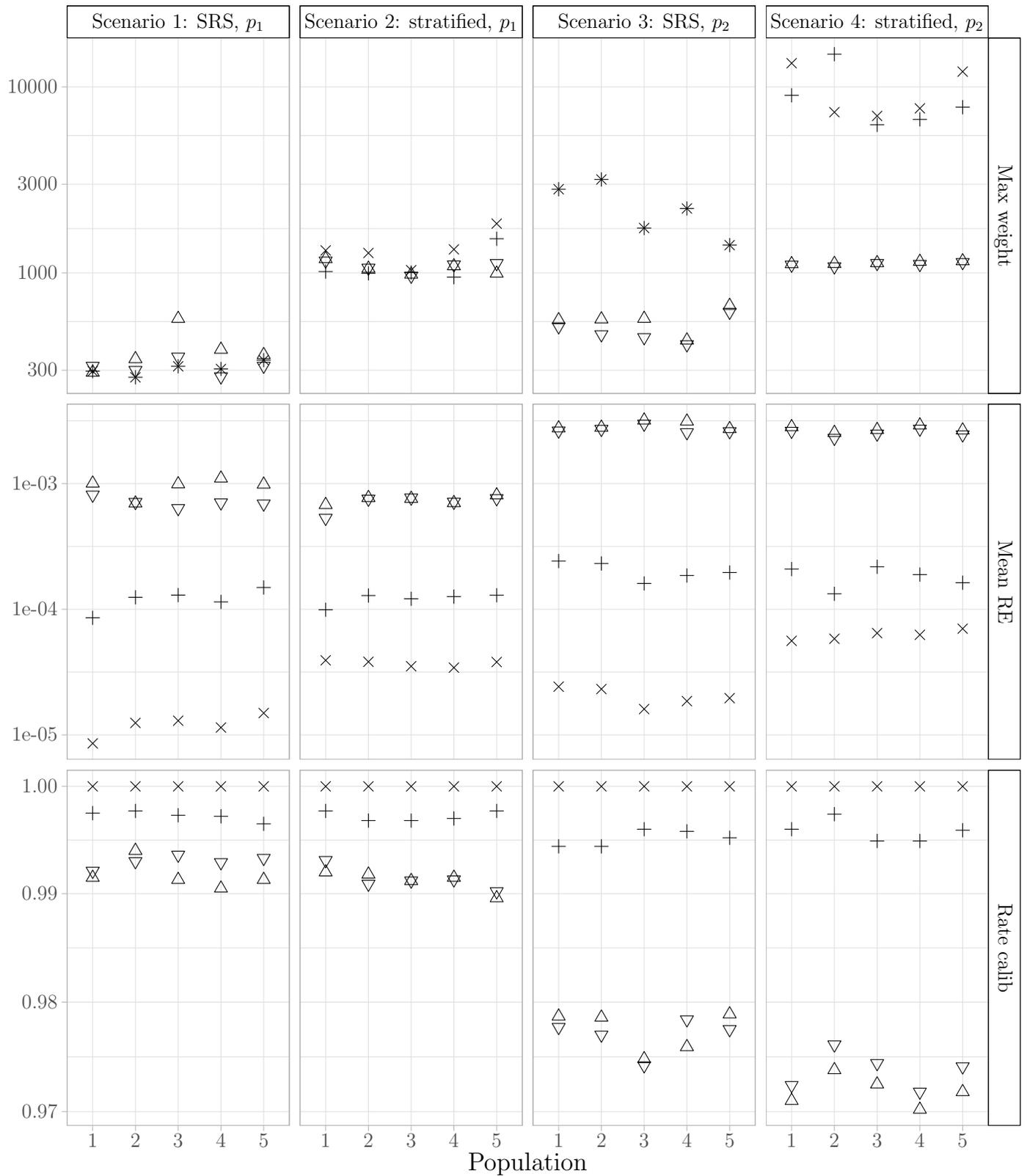}
\caption{Max weight, mean relative error, and rate of calibration for four NWA estimators, five populations, and 4 scenarios.\label{fig:convergence}}
\end{figure}

The results are presented in Figure~\ref{fig:convergence}. The y-axes are displayed in logarithmic scales. One estimator yields a Max weight of more than 400,000 in Scenario 4. To ease the reading of the graphs, this value is set to 15,000. In scenarios 1 and 2, when the model for the response probabilities is correctly specified, all four NWA estimators yield max weights close to one another. No extreme weights is noticeable. In scenario 3 and 4, when the model for the response probabilities is incorrectly specified, very large weights are obtained with MLE, more so in Scenario 4. Calibration may protect against extreme weights when the response model is misspecified. In all four scenarios, the mean RE is smaller with MLE than with calibration. This difference is larger in scenarios 3 and 4, when the model for the response probabilities is incorrectly specified. Moreover, the algorithm yields a mean RE smaller than the threshold of 0.01 more often with MLE than with calibration. This illustrates how the algorithm applied to obtain the response model parameters converges more often to a solution to the estimating equations of MLE than to a solution to the estimating equations of calibration.

%When estimating the response probabilities via calibration at the population level, we attempt to correct both the sampling and nonresponse errors. This may seem attractive but there are two main pitfalls highlighted in the simulation study: 1) convergence problems and 2) extreme weights.
%
%
%First, depending on the sampling design and nonresponse mechanism, it may happen that the optimization program fails to converge to a solution to the estimating equations. In the simulations, it only happened when calibrating at the population and with Poisson sampling in 2 to 3\% of the simulations in all three populations. These figures may change if we change convergence tolerances. Second, it may happen that the optimizer returns final weights $1/(\pi_i \widehat{p}_i)$ very large compared to other weights or to the true unobservable weights $1/(\pi_i p_i)$.
%
%
%Table~\ref{table:weights} shows the maximum final weights across all simulations. The maximum final weights are all close to one another and to the true unobservable weights $1/(\pi_i p_i)$, except for the weights obtained via calibration at the population level for Poisson sampling.

%\input{figs/table_weights_clean.tex}
\newgeometry{left=0.6in,right=0.6in, top =0.6in, bottom =1in}
\thispagestyle{empty}
%\begin{landscape}
%\input{figs/table_tot_srs_pois_clean.tex}
%\input{figs/table_tot_srs.tex}
%\input{figs/table_tot_pois.tex}
%\end{landscape}
\restoregeometry

\section{Discussion}\label{section:discussion}
% ----------------------------------------------------------------

We build on \cite{kim:kim:07} and develop asymptotic properties of the NWA estimator when calibration is applied to estimate the response probabilities. For the first time, a common theoretical framework is considered for both approaches to NWA estimation, namely MLE and calibration. This allows us to compare the asymptotic behavior of four estimators in terms of bias and variance under common assumptions. We postulate a logistic regression model for the response probabilities. We consider two levels of calibration: population and full sample. The main results are 1) the NWA estimators with the response probabilities estimated via calibration are asymptotically equivalent to unbiased estimators, 2) a gain in efficiency is obtained when estimating the response probabilities via calibration as compared to the estimator with the true response probabilities, 3) the NWA estimator with the response probabilities estimated via calibration at the population level is generally more efficient than the NWA estimator with response probabilities estimated via calibration at the sample level, 4) calibration may better protect against model misspecification than maximum likelihood when applied to estimate the response probabilities, and 5) we explain and illustrate the problems of convergence to a solution to the estimating equations and extreme weights. The paper studies and compares NWA estimators obtained either via MLE or direct calibration (one-step approach). Some authors suggest the two-step approach, i.e., first estimate the response probabilities via MLE in order to bypass the problem of extreme weights and then calibrate to further improve the efficiency of the NWA estimator, see~\cite{haz:les:16} and~\cite{haz:bea:17:weights:review}, p.222. This goes beyond the scope of this research and is the subject of future work.

%% ----------------------------------------------------------------
\section{Acknowledgments}
%% ----------------------------------------------------------------
This research was supported by the Swiss Federal Statistical Office. The author thanks Pr. Yves Till\'e, two referees, an Associate Editor, and the Editor of the journal Survey Methodology for constructive comments. The views expressed in this article are those of the author solely and do not necessarily reflect these of the aforementioned organization and persons.

\clearpage

\section*{\centering Inference from Sampling with Response Probabilities Estimated via Calibration\\Appendix}%
%\author{Caren Hasler\footnote{Institute of Statistics, University of Neuch\^atel, Av. de Bellevaux 51, 2000 Neuch\^atel, Switzerland.}}
%~~and
%Radu V. Craiu\footnote{Department of Statistical Sciences, University of Toronto, 100 St. Georges Street, Toronto, Ontario, M5S 3G3, Canada}}
%\date{}

%\begin{document}

%\maketitle

This appendix contains the proof of Results~\ref{result:asymptotics:calS} to \ref{result:robustness:nr}. Results~\ref{result:asymptotics:calS} and \ref{result:asymptotics:calU} show that the NWA estimators with response probabilities estimated via calibration at the sample level, respectively population level, are asymptotically equivalent to linearized estimators. The linearized estimators are unbiased for the population total. Results~\ref{result:robustness:sup} and \ref{result:robustness:nr} state that the NWA estimators with response probabilities estimated via calibration are consistent even when one of the models, response model or superpopulation model, is missspecified. That is, these estimators are doubly robust. We first present additional required assumptions about the sequence of finite populations, then restate the results, and provide proofs.

\setcounter{result}{0}

The following assumptions about the sequence of finite populations are needed for the proof of the results.
\begin{enumerate}[({P}1):]
    \item \label{assumption:bounded:moments} The study variable $y_i$ has bounded second and forth moments, and $\xb_i$, $\xb_i y_i$, $\xb_i\xb_i^\top y_i$ have bounded first moment, i.e.,
            \begin{align}
                 \limsup\limits_{N\rightarrow + \infty} \frac{1}{N} \sum_{i \in U} u_i < +\infty,
            \end{align}
            where $u_i$ is the vector obtained by stacking all columns of $y_i^2$, $y_i^4$, $\xb_i$, $\xb_i y_i$, and $\xb_i\xb_i^\top y_i$.
    \item \label{assumption:xtx} The population first moment of $\xb_i\xb_i^\top$ has a limit which is an invertible matrix, i.e.,
             \begin{align}
                 \lim\limits_{N\rightarrow + \infty} \frac{1}{N} \sum_{i \in U} \xb_i\xb_i^\top = \Tb,
            \end{align}
            where $\Tb$ is invertible.
    \item \label{assumption:sample:moments}All components of the sample moments of $\xb_i$, $\xb_i y_i$, $\xb_i\xb_i^\top$, and $\xb_i\xb_i^\top y_i$ converge to their population moments
            \begin{align}
                 \frac{1}{N} \sum_{i \in S} \frac{v_i}{\pi_i}   - \frac{1}{N} \sum_{i \in U}v_i  = O_p\left(n^{-1/2}\right),
            \end{align}
            where $v_i$ is obtained by stacking all columns of $\xb_i$, $\xb_i y_i$, $\xb_i\xb_i^\top$, and $\xb_i\xb_i^\top y_i$.
\item \label{assumption:respondents:moments} All components of the respondents moments of $\xb_i$, $\xb_i y_i$, $\xb_i\xb_i^\top$, and $\xb_i\xb_i^\top y_i$ converge to their sample moments
            \begin{align}
                 \frac{1}{N} \sum_{i \in S_r} \frac{v_i}{\pi_i p_i}   - \frac{1}{N} \sum_{i \in S}\frac{v_i}{\pi_i}  = O_p\left(n^{-1/2}\right),
            \end{align}
            where $v_i$ is defined above.
%    \item \label{assumption:bounded:moments} \re{Do we need eq(9) of Kim and Kim in the proof?}
%        \begin{align}\label{eqn:bounded:moments}
%            p\lim_{N\rightarrow +\infty} \frac{1}{N}\sum_{i \in S}\frac{1}{\pi_i} \left[ \xb_i, \xb_i \xb_i^\top \right]y_i
%< + \infty        \end{align}
\item \label{assumption:exp:lambda:x} The estimators $\widehat{\lambdab}^{cal,S}$ and $\widehat{\lambdab}^{cal,U}$  satisfy
            \begin{align}
      \exp \left(  -\xb_i^\top \widehat{\lambdab}^{cal,S}\right) &= O_p(1),\\
      \exp \left(  -\xb_i^\top \widehat{\lambdab}^{cal,U}\right) &= O_p(1).
    \end{align}
    This avoids to have some infinite nonresponse weights, i.e., some infinite inverse estimated response probabilities.
\item \label{assumption:lambda} The estimators $\widehat{\lambdab}^{cal,S}$ and $\widehat{\lambdab}^{cal,U}$  satisfy
        \begin{align}
            \widehat{\lambdab}^{cal,S} - \lambdab^0 &= -\left[\E_q\left\{\left.\frac{\partial}{\partial \lambdab}Q^{cal,S}(\lambdab^0)\right|S\right\}\right]^{-1}Q^{cal,S}(\lambdab^0) \label{eqn:lambda:diffS}\\
                                    &+o_p(n^{-1/2}),\\
            \widehat{\lambdab}^{cal,U} - \lambdab^0 &= -\left[\E_p\E_q\left\{\left.\frac{\partial}{\partial \lambdab}Q^{cal,U}(\lambdab^0)\right| S\right\}\right]^{-1}Q^{cal,U}(\lambdab^0) \label{eqn:lambda:diffU}\\
                                    &+o_p(n^{-1/2}).
            \end{align}
        This condition is a Taylor expansion of $\E_q\left\{\left.Q^{cal,S}(\widehat{\lambdab}^{cal,S})\right|S\right\}$, respectively of $\E_p\E_q\left\{\left.Q^{cal,U}(\widehat{\lambdab}^{cal,S})\right|S\right\}$, around $\lambdab^0$ with a second order remainder that is $o_p(n^{-1/2})$, and where we used $Q^{cal,S}(\widehat{\lambdab}^{cal,S})= Q^{cal,U}(\widehat{\lambdab}^{cal,U})=0$.
\end{enumerate}

\begin{result}\label{result:asymptotics:calS:appendix}
Let the sequence of sampling designs satisfy Assumptions~(D\ref{assumption:f})-(D\ref{assumption:deltaij}), the response mechanism satisfy Assumptions~(R\ref{assumption:bounded:p})-(R\ref{assumption:logistic}), and the sequence of finite populations satisfy Assumptions (P\ref{assumption:bounded:moments})-(P\ref{assumption:lambda}). % on the response mechanism and the moments of the variables $y_i$ and $\xb_i$ hold.
The NWA estimator $\widehat{Y}_{\widehat{p}}^{cal,S}$ satisfies
    \begin{align}
  \frac{1}{N}\widehat{Y}_{\widehat{p}}^{cal,S} &= \frac{1}{N}\widehat{Y}_{\widehat{p},l}^{cal,S} + O_p(n^{-1}),
    \end{align}
where
    \begin{align}
  \widehat{Y}_{\widehat{p},l}^{cal,S} &= \sum_{i \in S} \frac{1}{\pi_i}\left\{  \xb_i^\top\gammab_{S} + \frac{r_i}{p_i}\left(y_i -  \xb_i^\top\gammab_{S}\right) \right\},\\
  \gammab_{S} &= \left( \sum_{i \in S} \frac{1-p_i}{\pi_i}\xb_i\xb_i^\top \right)^{-1}\sum_{i \in S} \frac{1-p_i}{\pi_i}\xb_iy_i.
    \end{align}
\end{result}

\begin{result}\label{result:asymptotics:calU:appendix}
Let the sequence of sampling designs satisfy Assumptions~(D\ref{assumption:f})-(D\ref{assumption:deltaij}), the response mechanism satisfy Assumptions~(R\ref{assumption:bounded:p})-(R\ref{assumption:logistic}), and the sequence of finite populations satisfy Assumptions (P\ref{assumption:bounded:moments})-(P\ref{assumption:lambda}). % on the response mechanism and the moments of the variables $y_i$ and $\xb_i$ hold.
The NWA estimator $\widehat{Y}_{\widehat{p}}^{cal,U}$ satisfies
    \begin{align}
  \frac{1}{N}\widehat{Y}_{\widehat{p}}^{cal,U} &= \frac{1}{N}\widehat{Y}_{\widehat{p},l}^{cal,U} + O_p(n^{-1}),
    \end{align}
where
    \begin{align}
  \widehat{Y}_{\widehat{p},l}^{cal,U} &=  \sum_{i \in U} \left\{  \xb_i^\top\gammab_{U} + \frac{a_i}{\pi_i}\frac{r_i}{p_i}\left(y_i -  \xb_i^\top\gammab_{U}\right)  \right\},\\
\gammab_{U} &= \left\{ \sum_{i \in U} (1-p_i)\xb_i\xb_i^\top \right\}^{-1} \sum_{i \in U} (1-p_i)\xb_i y_i,\\
    \end{align}
\end{result}

%\begin{result}
%  Under the assumptions on the response mechanism, the sequence of sampling designs and of finite populations, we have
%\begin{enumerate}
%  \item The true parameter vector $\lambdab^0$ satisfies
%        \begin{align}
%            \E_q\left(Q^{cal,S}\left(\lambdab^0\right)|S\right) &= 0\\
%            \E_p\left(\E_q\left(Q^{cal,U}\left(\lambdab^0\right)|S\right) \right) &= 0
%        \end{align}
%    \item \re{Check both these and check what is needed in the proof:}
%        \end{align}
%\end{enumerate}
%\end{result}
%
%\begin{proof}
%    \begin{enumerate}
%        \item The equalities follows directly by taking the corresponding expectation of the estimating equations~\eqref{eqn:calU} and \eqref{eqn:calS}.
%        \item \re{Follow from the previous equalities and ???}
%    \end{enumerate}
%\end{proof}

To prove Results~\ref{result:asymptotics:calS:appendix} and~\ref{result:asymptotics:calU:appendix}, we will need the following two Lemmas.

\begin{lemma}\label{lemma:p}
    Suppose that Assumption (R\ref{assumption:logistic}) holds. The response probabilities satisfy
    \begin{align}
        p(\xb_i;\lambdab)^{-1} &= 1 + \exp\left( -\xb_i^\top  \lambdab\right),\\
        \frac{\partial p(\xb_i;\lambdab)^{-1}}{\partial \lambdab} &= -\exp\left( -\xb_i^\top  \lambdab\right)\xb_i = \left\{1-p(\xb_i;\lambdab)^{-1}\right\}\xb_i,\\
        \frac{\partial^2 p(\xb_i;\lambdab)^{-1}}{\partial \lambdab\partial\lambdab^\top} &= \exp\left( -\xb_i^\top  \lambdab\right)\xb_i \xb_i^\top =  p(\xb_i;\lambdab)^{-1}\left\{1 - p(\xb_i;\lambdab) \right\}\xb_i.
    \end{align}
\end{lemma}

\begin{proof}
  Straightforward computations yield the results.
\end{proof}

\begin{lemma}\label{lemma:lambda}
  Suppose that Assumptions (R\ref{assumption:logistic}) and (P\ref{assumption:lambda}) hold. We have
    \begin{align}
      \widehat{\lambdab}^{cal,S} - \lambdab^0 &= \Tb_{p\pi}^{-1}\left( \sum_{i \in U}\frac{a_i}{\pi_i}\frac{r_i}{p_i}\xb_i -  \sum_{i \in U}\frac{a_i}{\pi_i}\xb_i \right) + o_p(n^{-1/2}),\label{eqn:lambda:diffS3}\\
      \widehat{\lambdab}^{cal,U} - \lambdab^0 &= \Tb_{p}^{-1}\left( \sum_{i \in U}\frac{a_i}{\pi_i}\frac{r_i}{p_i}\xb_i -  \sum_{i \in U}\xb_i \right) + o_p(n^{-1/2}),\label{eqn:lambda:diffU3}\\
    \end{align}
    where
    \begin{align}
      \Tb_{p\pi} &=  \sum_{i \in U}\frac{a_i}{\pi_i}(1-p_i)\xb_i \xb_i^\top,\\
      \Tb_{p} &=  \sum_{i \in U}(1-p_i)\xb_i \xb_i^\top.
    \end{align}
\end{lemma}

\begin{proof}
 From Lemma~\ref{lemma:p}, we have
  \begin{align}
    \frac{\partial}{\partial \lambdab}Q^{cal,S}(\lambdab^0) &= \sum_{i \in U}\frac{a_i}{\pi_i}r_i  \left\{1-p(\xb_i;\lambdab^0)^{-1}\right\} \xb_i\xb_i^\top.
  \end{align}
  Therefore,
  \begin{align}
    \E_q\left\{\left.\frac{\partial}{\partial \lambdab}Q^{cal,S}(\lambdab^0)\right|S\right\}  &= \sum_{i \in U}\frac{a_i}{\pi_i} p(\xb_i;\lambdab^0) \left\{1-p(\xb_i;\lambdab^0)^{-1}\right\} \xb_i\xb_i^\top\\
            &= - \sum_{i \in U}\frac{a_i}{\pi_i} \left\{1-p(\xb_i;\lambdab^0)\right\} \xb_i\xb_i^\top.
  \end{align}
  We plug in Equation~\eqref{eqn:lambda:diffS} and obtain \eqref{eqn:lambda:diffS3}. From Lemma~\ref{lemma:p}, we have
  \begin{align}
    \frac{\partial}{\partial \lambdab}Q^{cal,U}(\lambdab^0) &= \sum_{i \in U}\frac{a_i}{\pi_i}r_i  \left\{1-p(\xb_i;\lambdab^0)^{-1}\right\} \xb_i\xb_i^\top.
  \end{align}
  Therefore,
  \begin{align}
    \E_p\E_q\left\{\left.\frac{\partial}{\partial \lambdab}Q^{cal,U}(\lambdab^0)\right|S\right\}  &= \sum_{i \in U} p(\xb_i;\lambdab^0) \left\{1-p(\xb_i;\lambdab^0)^{-1}\right\} \xb_i\xb_i^\top\\
            &= - \sum_{i \in U}\left\{1-p(\xb_i;\lambdab^0)\right\} \xb_i\xb_i^\top.
  \end{align}
  We plug in Equation~\eqref{eqn:lambda:diffU} and obtain \eqref{eqn:lambda:diffU3}.
\end{proof}

%\begin{lemma}
%    \re{Let the sequence of sampling designs satisfy Assumptions~(D\ref{assumption:f})-(D\ref{assumption:deltaij}), the response mechanism satisfy Assumptions~(R\ref{assumption:MAR})-(R\ref{assumption:logistic}), and the sequence of finite populations satisfy Assumptions (P\ref{assumption:bounded:moments})-(P\ref{assumption:lambda}). We have
%    \begin{align}
%      \exp \left(  \xb_i^\top \widehat{\lambdab}^{cal,S}\right) &= O_p\left[exp(n^{-1})\right],\\
%      \exp \left(  \xb_i^\top \widehat{\lambdab}^{cal,U}\right) &= O_p\left[exp(n^{-1})\right].
%    \end{align}}
%\end{lemma}
%
%\begin{proof}
%  We can easily show that $\frac{F'( \xb_i^\top\lambdab^0)}{F( \xb_i^\top\lambdab^0)} \in \left]-1;0\right[$. Therefore, from Assumption~(P\ref{assumption:xtx}) we obtain $N\Tb_{F}^{-1} = O(1)$. Moreover, from Assumption~(D\ref{assumption:pii:bounded}) we obtain $N\Tb_{F\pi}^{-1} = O(1)$. Equation\eqref{eqn:lambda:diffS3} can be rewritten
%  \begin{align}
%    \widehat{\lambdab}^{cal,S}  &=\lambdab^0 +  N\Tb_{F\pi}^{-1}\left(\frac{1}{N} \sum_{i \in U}\frac{a_i}{\pi_i}\frac{r_i}{p_i}\xb_i -  \frac{1}{N}\sum_{i \in U}\frac{a_i}{\pi_i}\xb_i \right) + o_p(n^{-1/2}),
%  \end{align}
%  Using Assumption~(P\ref{assumption:respondents:moments}) and $N\Tb_{F}^{-1} = O(1)$, we obtain $\widehat{\lambdab}^{cal,S}= \lambdab^0 + O_p(n^{-1/2})$. Similarly, we can show that $\widehat{\lambdab}^{cal,U}=  \lambdab^0 + O_p(n^{-1/2})$.
%\end{proof}

\begin{proof}[Proof of Results~\ref{result:asymptotics:calS:appendix} and \ref{result:asymptotics:calU:appendix}]
  %The first part of the proof is similar to the first part of the proof of Theorem 1 of~\cite{kim:kim:07}. The second part, from Equation~\eqref{eqn:lambda:diffU2}, is specific to the NWA estimators obtained via calibration.
  When calibration at the population level is used, a second-order Taylor development of the estimated response probabilities is
\begin{align}
  p\left(\xb_i;\widehat{\lambdab}^{cal,U}\right)^{-1} &= p(\xb_i;\lambdab^0)^{-1} + \left\{\left.\frac{\partial p(\xb_i;\lambdab)^{-1}}{\partial \lambdab} \right|_{\lambdab = \lambdab^0}\right\}^\top \left(\widehat{\lambdab}^{cal,U} - \lambdab^0\right)\\
    &+ \frac{1}{2} \left(\widehat{\lambdab}^{cal,U} - \lambdab^0\right)^\top\left\{\left.\frac{\partial^2 p(\xb_i;\lambdab)^{-1}}{\partial \lambdab\partial\lambdab^\top} \right|_{\lambdab = \tilde{\lambdab}}\right\} \left(\widehat{\lambdab}^{cal,U} - \lambdab^0\right),
\end{align}
where $\tilde{\lambdab}$ is on the line segment between $\widehat{\lambdab}^{cal,U}$ and $\lambdab^0$. The NWA estimator $\widehat{Y}_{\widehat{p}}^{cal,U}$ can be written as
\begin{align}\label{eqn:nwa:calU:asym}
 \frac{1}{N}\widehat{Y}_{\widehat{p},l}^{cal,U} &=  \frac{1}{N} \sum_{i \in S}\frac{r_i}{\pi_i}p\left(\xb_i;\widehat{\lambdab}^{cal,U}\right)^{-1}  y_i\\
        &=\frac{1}{N} \sum_{i \in S}\frac{r_i}{\pi_i}p(\xb_i;\lambdab^0)^{-1}  y_i + A_N^\top \left(\widehat{\lambdab}^{cal,U} - \lambdab^0\right)\\
        &+ \frac{1}{2} \left(\widehat{\lambdab}^{cal,U} - \lambdab^0\right)^\top B_N \left(\widehat{\lambdab}^{cal,U} - \lambdab^0\right),
\end{align}
where
\begin{align}
  A_N &= \frac{1}{N}\sum_{i \in S}\frac{r_i}{\pi_i} \left\{\left.\frac{\partial p(\xb_i;\lambdab)^{-1}}{\partial \lambdab} \right|_{\lambdab = \lambdab^0}\right\}y_i,\\
B_N &= \frac{1}{N}\sum_{i \in S}\frac{r_i}{\pi_i}\left\{\left.\frac{\partial^2 p(\xb_i;\lambdab)^{-1}}{\partial \lambdab\partial\lambdab^\top} \right|_{\lambdab = \tilde{\lambdab}}\right\} y_i.
\end{align}
Lemma~\ref{lemma:p} yields
\begin{align}
  A_N &= \frac{1}{N}\sum_{i \in S}\frac{r_i}{\pi_i}\frac{p_i-1}{p_i} \xb_i y_i,\\
B_N &= \frac{1}{N}\sum_{i \in S}\frac{r_i}{\pi_i} \frac{1 - p(\xb_i;\tilde{\lambdab})}{p(\xb_i;\tilde{\lambdab})}\xb_i \xb_i^\top y_i,
\end{align}
where we used $p_i = p(\xb_i;\lambdab^0)$.
%\re{where $\tilde{p}_i$ is $p(\xb_i;\lambdab)$ evaluated at $\tilde{\lambdab}$}.
From assumption~(P\ref{assumption:respondents:moments}), we have
\begin{align}\label{eqn:AN}
  A_N &= \frac{1}{N}\sum_{i \in S}\frac{p_i-1}{\pi_i} \xb_i y_i + O_p(n^{-1/2}),
\end{align}
and from assumption~(P\ref{assumption:sample:moments}) we obtain
\begin{align}\label{eqn:AN:U}
  A_N &= \frac{1}{N}\sum_{i \in U} (p_i-1)\xb_i y_i + O_p(n^{-1/2}).
\end{align}
As for $B_N$, we can write using (P\ref{assumption:sample:moments}) and (P\ref{assumption:respondents:moments})
\begin{align}
B_N &= \frac{1}{N}\sum_{i \in U} p_i \frac{1 - p(\xb_i;\tilde{\lambdab})}{p(\xb_i;\tilde{\lambdab})}\xb_i \xb_i^\top y_i + O_p(n^{-1/2}).
\end{align}
Since $p(\xb_i;\lambdab) = \left[1+ \exp(-\xb_i^\top \lambdab)\right]$, $B_N$ can be written
\begin{align}
B_N &= \frac{1}{N}\sum_{i \in U} p_i \exp(-\xb_i^\top \tilde{\lambdab})\xb_i \xb_i^\top y_i + O_p(n^{-1/2}).
\end{align}
From Assumptions~(R\ref{assumption:bounded:p}), we have $\exp(-\xb_i^\top \lambdab^0)=O(1)$. Moreover, from Assumption (P\ref{assumption:exp:lambda:x}) and since $\tilde{\lambdab}$ is on the line segment between $\widehat{\lambdab}^{cal,U}$ and $\lambdab^0$, we have $\exp(-\xb_i^\top \tilde{\lambdab}) = O_p(1)$. Using this result and assumption~(P\ref{assumption:bounded:moments}), we obtain
\begin{align}\label{eqn:BN}
  B_N &= O_p(1).
\end{align}
%From Equation~\eqref{eqn:lambda:diffU}, it comes
%\begin{align}\label{eqn:lambda:diffU2}
%    \widehat{\lambdab}^{cal,U} - \lambdab^0 &=
%\left\{ \sum_{i \in U} (1-p_i)\xb_i\xb_i^\top \right\}^{-1}\sum_{i \in U} \left( \frac{a_i}{\pi_i}\frac{r_i}{p_i} - 1 \right)\xb_i+o_p(n^{-1/2}).
%\end{align}
Inserting Equations~\eqref{eqn:lambda:diffU3},~\eqref{eqn:AN:U}, and~\eqref{eqn:BN} into Equation~\eqref{eqn:nwa:calU:asym} yields
\begin{align}
    \frac{1}{N}\widehat{Y}_{\widehat{p}}^{cal,U} &= \frac{1}{N} \sum_{i \in U}\left\{ \xb_i^\top\gammab_{U} +   \frac{a_i}{\pi_i}\frac{r_i}{p_i} \left( y_i - \xb_i^\top \gammab_{U} \right)\right\}     + O_p(n^{-1}),
\end{align}
where
\begin{align}
  \gammab_{U} &= \left\{ \sum_{i \in U} (1-p_i)\xb_i\xb_i^\top \right\}^{-1} \sum_{i \in U} (1-p_i)\xb_i y_i.
\end{align}
A similar construction for $\widehat{Y}_{\widehat{p}}^{cal,S}$ yields
\begin{align}\label{eqn:nwa:calS:asym}
 \frac{1}{N}\widehat{Y}_{\widehat{p}}^{cal,S} &= \frac{1}{N} \sum_{i \in S}\frac{1}{\pi_i}\frac{r_i}{p_i}y_i + A_N^\top \left(\widehat{\lambdab}^{cal,S} - \lambdab^0\right)\\
 &+ \frac{1}{2} \left(\widehat{\lambdab}^{cal,S} - \lambdab^0\right)^\top B_N \left(\widehat{\lambdab}^{cal,S} - \lambdab^0\right),
\end{align}
where $A_N$ and $B_N$ are as above. Inserting Equations~\eqref{eqn:lambda:diffS3},~\eqref{eqn:AN}, and~\eqref{eqn:BN} into Equation~\eqref{eqn:nwa:calS:asym} yields
\begin{align}
    \frac{1}{N}\widehat{Y}_{\widehat{p}}^{cal,S} &= \frac{1}{N} \sum_{i \in S}\frac{1}{\pi_i}\left\{ \xb_i^\top\gammab_{S} +   \frac{r_i}{p_i} \left( y_i - \xb_i^\top \gammab_{S} \right)\right\}     + O_p(n^{-1}),
\end{align}
where
\begin{align}
  \gammab_{S} &= \left( \sum_{i \in S} \frac{1-p_i}{\pi_i} \xb_i\xb_i^\top \right)^{-1} \sum_{i \in S}\frac{1-p_i}{\pi_i} \xb_i y_i.
\end{align}
\end{proof}

\begin{result}
  Consider the superpopulation model $\xi: y_i = \xb_i^\top \betab + \varepsilon_i$ where $\E_{\xi}(\varepsilon_i) = 0$, $\E_{\xi}(\varepsilon_i\varepsilon_j) = \sigma^2 \leq +\infty$ if $i = j$ and 0 otherwise, and subscript $\xi$ means that the expectation and variance are computed with respect to model $\xi$. Suppose that assumptions (D\ref{assumption:f})-(D\ref{assumption:deltaij}), (R\ref{assumption:bounded:p}), (R\ref{assumption:MAR}), (R\ref{assumption:bounded:phat}) are satisfied. Then
  \begin{align}
    \frac{\widehat{Y}_{\widehat{p}}^{cal,U} - Y}{N} &= o_{\mathbb{P}}(1),\\
    \frac{\widehat{Y}_{\widehat{p}}^{cal,S} - Y}{N} &= o_{\mathbb{P}}(1).
  \end{align}
  Subscript $\mathbb{P}$ means that the reference probability distribution is that determined by the superpopulation model, the sampling design, and the response mechanism.
\end{result}

\begin{proof}
  For the first part, we show that $\E_{\xi p q}\left( \frac{\widehat{Y}_{\widehat{p}}^{cal,U} - Y}{N} \right) = 0$ and that \newline$\V_{\xi p q}\left( \frac{\widehat{Y}_{\widehat{p}}^{cal,U} - Y}{N} \right)$ converges to 0. From Assumption~(R\ref{assumption:MAR}) and since the sampling design is non-informative, we can switch the expectation with respect to $\xi$ with the expectation with respect to $p$ and $q$. We can write
  \begin{align}
    \E_{\xi p q}\left( \frac{\widehat{Y}_{\widehat{p}}^{cal,U} - Y}{N} \right) &= \frac{1}{N}\E_{p q \xi} \left( \widehat{Y}_{\widehat{p}}^{cal,U} - Y \right)\\
                       &=  \frac{1}{N}\betab^\top \E_{p q } \left( \sum_{i \in S_r}\frac{\xb_i}{\pi_i \widehat{p}_i} - \sum_{i \in U}\xb_i \right)\\
                       &=  \frac{1}{N}\betab^\top \E_{p q } \left( \sum_{i \in U}\xb_i - \sum_{i \in U}\xb_i \right) = 0
  \end{align}
  where the second equality comes from the superpopulation model and the last one from estimating Equation~\eqref{eqn:calU:S}. As far as the variance is concerned,
  \begin{align}
    \V_{\xi p q}\left( \frac{\widehat{Y}_{\widehat{p}}^{cal,U} - Y}{N} \right) &= \frac{1}{N^2}\V_{\xi p q} \left( \widehat{Y}_{\widehat{p}}^{cal,U} \right)\\
                       &=  \frac{1}{N^2} \V_{\xi p q } \left( \sum_{i \in S_r}\frac{\betab^\top\xb_i + \varepsilon_i}{\pi_i \widehat{p}_i} \right)\\
                       &=  \frac{1}{N^2} \V_{\xi p q } \left( \betab^\top \sum_{i \in U}\xb_i + \sum_{i \in S_r}\frac{\varepsilon_i}{\pi_i \widehat{p}_i}\right)
  \end{align}
  where the last equality comes from estimating Equation~\eqref{eqn:calU:S}. From Assumption~(R\ref{assumption:MAR}) and since the sampling design is non-informative we can write
    \begin{align}
    \V_{\xi p q}\left( \frac{\widehat{Y}_{\widehat{p}}^{cal,U} - Y}{N} \right) &= \frac{1}{N^2} \V_{\xi p q } \left( \sum_{i \in S_r}\frac{\varepsilon_i}{\pi_i \widehat{p}_i}\right)\\
            &= \frac{1}{N^2} \left[\V_{\xi}\E_{ p q } \left( \sum_{i \in S_r}\frac{\varepsilon_i}{\pi_i \widehat{p}_i}\right) + \E_{\xi}\V_{ p q } \left( \sum_{i \in S_r}\frac{\varepsilon_i}{\pi_i \widehat{p}_i}\right)\right] \\
            &= \frac{1}{N^2} \left[\E_{ p q }\V_{\xi} \left( \sum_{i \in S_r}\frac{\varepsilon_i}{\pi_i \widehat{p}_i}\right) + \V_{ p q } \E_{\xi}\left( \sum_{i \in S_r}\frac{\varepsilon_i}{\pi_i \widehat{p}_i}\right)\right].
  \end{align}
  Using that $\E_{\xi}(\varepsilon_i) = 0$ and $\E_{\xi}(\varepsilon_i\varepsilon_j) = \sigma^2 \leq +\infty$ if $i = j$ and 0 otherwise, we obtain
  \begin{align}
    \V_{\xi p q}\left( \frac{\widehat{Y}_{\widehat{p}}^{cal,U} - Y}{N} \right)
            &= \frac{\sigma^2}{N^2} \E_{ p q }\left( \sum_{i \in S_r}\frac{1}{\pi_i^2 \widehat{p}_i^2}\right).
  \end{align}
  From Assumptions~(D\ref{assumption:pii:bounded}) and (R\ref{assumption:bounded:phat}), we have
    \begin{align}
    \V_{\xi p q}\left( \frac{\widehat{Y}_{\widehat{p}}^{cal,U} - Y}{N} \right)
            &\leq \frac{\sigma^2}{N^2}\frac{1}{c_1\lambda_1} \E_{ p q }\left( \sum_{i \in S_r}\frac{1}{\pi_i \widehat{p}_i}\right).
  \end{align}
  Using estimating Equation~\eqref{eqn:calU:S} and since the $\xb_i$'s contain a constant, we obtain
      \begin{align}
    \V_{\xi p q}\left( \frac{\widehat{Y}_{\widehat{p}}^{cal,U} - Y}{N} \right)
            &\leq \frac{\sigma^2}{N^2}\frac{1}{c_1\lambda_1} \E_{ p q }\left( \sum_{i \in U}1\right)\\
            &= \frac{\sigma^2}{N}\frac{1}{c_1\lambda_1}.
  \end{align}
  As a result, $\V_{\xi p q}\left( \frac{\widehat{Y}_{\widehat{p}}^{cal,U} - Y}{N}\right)$ converges to 0. We conclude that $\frac{\widehat{Y}_{\widehat{p}}^{cal,U} - Y}{N}$ converges to 0 in probability.

  For the second part, we show that $\E_{\xi p q}\left( \frac{\widehat{Y}_{\widehat{p}}^{cal,S} - Y}{N} \right) = 0$ and that $\V_{\xi p q}\left( \frac{\widehat{Y}_{\widehat{p}}^{cal,S} - Y}{N} \right)$ converges to 0. From Assumption~(R\ref{assumption:MAR}) and since the sampling design is non-informative, we can switch the expectation with respect to $\xi$ with the expectation with respect to $p$ and $q$. We can write
  \begin{align}
    \E_{\xi p q}\left( \frac{\widehat{Y}_{\widehat{p}}^{cal,S} - Y}{N} \right) &= \frac{1}{N}\E_{p q \xi} \left( \widehat{Y}_{\widehat{p}}^{cal,S} - Y \right)\\
                       &=  \frac{1}{N}\betab^\top \E_{p q } \left( \sum_{i \in S_r}\frac{\xb_i}{\pi_i \widehat{p}_i} - \sum_{i \in U}\xb_i \right)\\
                       &=  \frac{1}{N}\betab^\top \E_{p q } \left( \sum_{i \in S}\frac{\xb_i}{\pi_i} - \sum_{i \in U}\xb_i \right) = 0
  \end{align}
  where the second equality comes from the superpopulation model, the third one from estimating Equation~\eqref{eqn:calS:S}, and the last one from $\E_p\left(\sum_{i \in S}\frac{\xb_i}{\pi_i}\right) = \sum_{i \in U}\xb_i$. As far as the variance is concerned,
  \begin{align}
    \V_{\xi p q}\left( \frac{\widehat{Y}_{\widehat{p}}^{cal,S} - Y}{N} \right) &= \frac{1}{N^2}\V_{\xi p q} \left( \widehat{Y}_{\widehat{p}}^{cal,S} \right)\\
                       &=  \frac{1}{N^2} \V_{\xi p q } \left( \sum_{i \in S_r}\frac{\betab^\top\xb_i + \varepsilon_i}{\pi_i \widehat{p}_i} \right)\\
                       &=  \frac{1}{N^2} \V_{\xi p q } \left( \betab^\top \sum_{i \in S}\frac{\xb_i}{\pi_i} + \sum_{i \in S_r}\frac{\varepsilon_i}{\pi_i \widehat{p}_i}\right)\\
                       &=  \frac{1}{N^2} \V_{\xi p q } \left( \sum_{i \in S_r}\frac{\varepsilon_i}{\pi_i \widehat{p}_i}\right)
  \end{align}
  where the third equality comes from estimating Equation~\eqref{eqn:calS:S}. Similar computations as for $\widehat{Y}_{\widehat{p}}^{cal,U}$ yield
  \begin{align}
    \V_{\xi p q}\left( \frac{\widehat{Y}_{\widehat{p}}^{cal,S} - Y}{N} \right)
            &\leq \frac{\sigma^2}{N^2}\frac{1}{c_1\lambda_1} \E_{ p q }\left( \sum_{i \in S_r}\frac{1}{\pi_i \widehat{p}_i}\right).
  \end{align}
  Using estimating Equation~\eqref{eqn:calS:S} and since the $\xb_i$'s contain a constant, we obtain
      \begin{align}
    \V_{\xi p q}\left( \frac{\widehat{Y}_{\widehat{p}}^{cal,S} - Y}{N} \right)
            &\leq \frac{\sigma^2}{N^2}\frac{1}{c_1\lambda_1} \E_{ p q }\left( \sum_{i \in S}\frac{1}{\pi_i}\right)\\
            &= \frac{\sigma^2}{N^2}\frac{1}{c_1\lambda_1}\sum_{i \in U}1\\
            &= \frac{\sigma^2}{N}\frac{1}{c_1\lambda_1}.
  \end{align}
  As a result, $\V_{\xi p q}\left( \frac{\widehat{Y}_{\widehat{p}}^{cal,S} - Y}{N}\right)$ converges to 0. We conclude that $\frac{\widehat{Y}_{\widehat{p}}^{cal,S} - Y}{N}$ converges to 0 in probability.
\end{proof}

\begin{result}\label{result:robustness:nr:appendix}
  Let the sequence of sampling designs satisfy Assumptions~(D\ref{assumption:f})-(D\ref{assumption:deltaij}), the response mechanism satisfy Assumptions~(R\ref{assumption:bounded:p})-(R\ref{assumption:logistic}), and the sequence of finite populations satisfy Assumptions (P\ref{assumption:bounded:moments})-(P\ref{assumption:lambda}). Then
  \begin{align}
    \frac{\widehat{Y}_{\widehat{p}}^{cal,U} - Y}{N} &= o_{p}(1),\\
    \frac{\widehat{Y}_{\widehat{p}}^{cal,S} - Y}{N} &= o_{p}(1).
  \end{align}
\end{result}

For the proof of Result~\ref{result:robustness:nr:appendix}, we will need the following Lemma.
\begin{lemma}\label{lemma:gamma}
  Suppose that Assumptions~(P\ref{assumption:bounded:moments})-(P\ref{assumption:sample:moments}) are verified. Then $\gammab_S - \gammab_U$ is $o_p(1)$.
\end{lemma}

\begin{proof}
We can write
\begin{align}
  \gammab_S &= \left(\frac{1}{N}\Tb_{p\pi}\right)^{-1}\frac{1}{N}\tb_{p\pi},\\
  \gammab_U &= \left(\frac{1}{N}\Tb_{p}\right)^{-1}\frac{1}{N}\tb_{p},
\end{align}
where $\Tb_{p\pi}$ and $\Tb_{p}$ are defined in Lemma~\ref{lemma:lambda} and
\begin{align}
  \tb_{p\pi} &= \sum_{i \in U}\frac{a_i}{\pi_i}(1-p_i)\xb_i y_i,\\
  \tb_{p} &= \sum_{i \in U}(1-p_i)\xb_i y_i.
\end{align}
 We can write $\gammab_S - \gammab_U = C_N + D_N$, where
  \begin{align}
    C_N &= \left( \left[\left\{ \Ib_v +   \left(\frac{1}{N}\Tb_p\right)^{-1}\frac{1}{N}\Tb_{\varepsilon}\right\}^{-1} - \Ib_v \right]  \left(\frac{1}{N}\Tb_p\right)^{-1}  \right)\frac{1}{N}\tb_p,\\
    D_N &= \left[\left\{ \Ib_v +   \left(\frac{1}{N}\Tb_p\right)^{-1}\frac{1}{N}\Tb_{\varepsilon}\right\}^{-1}\left(\frac{1}{N}\Tb_p\right)^{-1} \right]\frac{1}{N}\tb_{\varepsilon},\\
    \Tb_{\varepsilon}  &= \Tb_{p\pi} - \Tb_{p},\\
    \tb_{\varepsilon} &= \tb_{p\pi} - \tb_{p},
  \end{align}
  and $\Ib_v$ is the identity matrix of order $v$. From Assumption~(P\ref{assumption:sample:moments}), $\frac{1}{N}\Tb_{\varepsilon}$ is $O_p(n^{-1/2})$. From Assumption~(P\ref{assumption:xtx}), $\left(\frac{1}{N}\Tb_p\right)^{-1}$ is $O(1)$. As a result,\newline $\left\{ \Ib_v +   \left(\frac{1}{N}\Tb_p\right)^{-1}\frac{1}{N}\Tb_{\varepsilon}\right\}^{-1} $ converges in probability to $\Ib_v$. Moreover, Assumption~(P\ref{assumption:bounded:moments}) states that $\frac{1}{N}\tb_p$ is $O(1)$. We can conclude that $C_N$ is $o_p(1)$. From Assumption~(P\ref{assumption:sample:moments}), we know that $\frac{1}{N}\tb_{\varepsilon}$ is $O_p(n^{-1/2})$. Together with the aforementioned asymptotic behavior of the other terms in $D_N$, we conclude that $D_N$ is $o_p(1)$. Finally, $\gammab_S - \gammab_U$ is $o_p(1)$ as sum of two terms that are $o_p(1)$.
\end{proof}

\begin{proof}[Proof of Result~\ref{result:robustness:nr:appendix}]
  For the first part, we use the decomposition
   \begin{align}
    \frac{\widehat{Y}_{\widehat{p}}^{cal,U} - Y}{N} &= \frac{\widehat{Y}_{\widehat{p}}^{cal,U} - \widehat{Y}_{\widehat{p},l}^{cal,U}}{N} + \frac{\widehat{Y}_{\widehat{p},l}^{cal,U} - Y}{N}
  \end{align}
  and show that each term is $o_{p}(1)$. From Result~\ref{result:asymptotics:calU:appendix},
  \begin{align}
    \frac{\widehat{Y}_{\widehat{p}}^{cal,U} -  \widehat{Y}_{\widehat{p},l}^{cal,U}}{N} &= o_{p}(1).
  \end{align}
  We now proceed to show that
  \begin{align}
    \frac{\widehat{Y}_{\widehat{p},l}^{cal,U} - Y}{N} &= o_{p}(1).
  \end{align}
  We have
  \begin{align}
    \E_{pq} \left( \frac{\widehat{Y}_{\widehat{p},l}^{cal,U} - Y}{N} \right) = 0.
  \end{align}
  Moreover,
  \begin{align}
    \V_{pq} &\left( \frac{\widehat{Y}_{\widehat{p},l}^{cal,U} - Y}{N} \right) = \frac{1}{N^2} \V_{pq} \left( \widehat{Y}_{\widehat{p},l}^{cal,U} \right)\\
            &= \frac{1}{N^2} \left[ \V_p\left\{ \sum_{i \in S} \frac{1}{\pi_i} \left(y_i -  \xb_i^\top\gammab_{U}\right) \right\} +  \E_p \left\{  \sum_{i \in S} \frac{1}{\pi_i^2}\frac{1 - p_i}{p_i} \left(y_i -  \xb_i^\top\gammab_{U}\right)^2 \right\} \right]\\
            &= \frac{1}{N^2}\left\{
            \sum_{i\in U} \frac{1-\pi_i}{\pi_i} \left(y_i - \xb_i^\top\gammab_{U}\right)^2 \right. \\
            &\qquad + \left. \sum_{i \in U}\sum_{j \in U; j\neq i} \frac{\pi_{ij} - \pi_i\pi_j}{\pi_i\pi_j}\left(y_i -  \xb_i^\top\gammab_{U}\right)\left(y_j -  \xb_j^\top\gammab_{U}\right) \right\}\\
            &\qquad + \left. \sum_{i \in U} \frac{1}{\pi_i}\frac{1-p_i}{p_i}\left(y_i -  \xb_i^\top\gammab_{U}\right)^2
            \right\}.
  \end{align}
 where the second equality comes from \eqref{eqn:Vsam:calU} and \eqref{eqn:Vnr:calU}. From Assumptions~(D\ref{assumption:pii:bounded}) and (R\ref{assumption:bounded:p}), we can majorate this quantity as follows
 \begin{align}
   \V_{pq}& \left( \frac{\widehat{Y}_{\widehat{p},l}^{cal,U} - Y}{N} \right)
        \leq   \frac{1}{N\lambda_1} \sum_{i\in U} \frac{ \left(y_i - \xb_i^\top\gammab_{U}\right)^2 }{N} \\
        &+ \frac{ \max\limits_{i,j\in U,i \neq j} \left|\pi_{ij} - \pi_i \pi_j \right|}{\lambda_1^2}\left(\sum_{i\in U} \frac{ \left|y_i - \xb_i^\top\gammab_{U}\right|}{N}\right)^2 \\
        &+ \frac{1}{N\lambda_1}\frac{1-c}{c}\sum_{i \in U} \frac{\left(y_i -  \xb_i^\top\gammab_{U}\right)^2}{N}
 \end{align}
 This quantity converges to 0 from Assumptions~(P\ref{assumption:bounded:moments}), (P\ref{assumption:xtx}), and (D\ref{assumption:deltaij}).

 For the second part, we use the following decomposition
 \begin{align}
    \frac{\widehat{Y}_{\widehat{p}}^{cal,S} - Y}{N} &=  \frac{\widehat{Y}_{\widehat{p}}^{cal,S} - \widehat{Y}_{\widehat{p},l}^{cal,S}}{N} + \frac{\widehat{Y}_{\widehat{p},l}^{cal,S} - \widehat{Y}_{\widehat{p},l}^{cal,U}}{N} + \frac{\widehat{Y}_{\widehat{p},l}^{cal,U} - Y}{N}
 \end{align}
 and show that each term is $o_{p}(1)$. From Result~\ref{result:asymptotics:calS}, the first term is $o_{p}(1)$. We have shown above that the third term is $o_{p}(1)$. We now proceed to show that the second term is $o_{p}(1)$. This term can be written
  \begin{align}
    \frac{\widehat{Y}_{\widehat{p},l}^{cal,S} - \widehat{Y}_{\widehat{p},l}^{cal,U}}{N} &=
        \frac{1}{N}\left\{ \sum_{i \in U}\frac{a_i}{\pi_i}\left(1 - \frac{r_i}{p_i}\right)\xb_i^\top\gammab_S - \sum_{i \in U}\left(1 - \frac{a_i}{\pi_i}\frac{r_i}{p_i}\right)\xb_i^\top\gammab_U  \right\}
 \end{align}
 Rearranging further yields $\frac{\widehat{Y}_{\widehat{p},l}^{cal,S} - \widehat{Y}_{\widehat{p},l}^{cal,U}}{N} = E_N + F_N$, where
 \begin{align}
    E_N &= \left(\gammab_S - \gammab_U\right)^\top \left(\frac{1}{N} \sum_{i \in U}\frac{a_i}{\pi_i}\xb_i - \frac{1}{N} \sum_{i \in U}\frac{a_i}{\pi_i}\frac{r_i}{p_i}\xb_i \right)\\
    F_N &= \gammab_U^\top \left(\frac{1}{N} \sum_{i \in U}\frac{a_i}{\pi_i}\xb_i - \frac{1}{N} \sum_{i \in U}\xb_i \right).
 \end{align}
 From Lemma~\ref{lemma:gamma}, we know that $\gammab_S - \gammab_U$ is $o_p(1)$. From Assumption~(P\ref{assumption:respondents:moments}), the second term in $E_N$ is $O_p(n^{-1/2})$. As a result, $E_N$ is $o_p(1)$. From Assumptions~(P\ref{assumption:bounded:moments}) and (P\ref{assumption:xtx}), we know that $\gammab_U$ is $O(1)$. From Assumption~(P\ref{assumption:sample:moments}), the second term in $F_N$ is $O_p(n^{-1/2})$. As a result, $F_N$ is $o_p(1)$. We can conclude that $\frac{\widehat{Y}_{\widehat{p},l}^{cal,S} - \widehat{Y}_{\widehat{p},l}^{cal,U}}{N}$ is $o_p(1)$ as sum of two terms that are $o_p(1)$.
\end{proof}

%\begin{thebibliography}{}
%
%\bibitem[Kim and Kim, 2007]{kim:kim:07}
%Kim, J.~K. and Kim, J. (2007).
%\newblock Nonresponse weighting adjustment using estimated response
%  probability.
%\newblock {\em The Canadian Journal of Statistics / La Revue Canadienne de
%  Statistique}, 35(4):501--514.
%
%
%\end{thebibliography}

%\bibliography{M:/Research/bib} % Caren's bib
%\bibliography{Z:/Bibliographie/bibyves} %Yves' bib
%\bibliography{bibyves}
%\bibliographystyle{apalike}

%\end{document} 
%\putbib[M:/Research/bib] % for use with bibunits
%\end{bibunit}

\clearpage

% ----------------------------------------------------------------

% ----------------------------------------------------------------
%\bibliography{M:/Research/bib} % Caren's bib
%%\bibliography{Z:/Bibliographie/bibyves} %Yves' bib

%\bibliography{C:/Users/haslerc2/switchdrive/Research/bib}
%\bibliography{C:/Users/Michael/switchdrive/Research/bib} % home
\bibliographystyle{chicago}

%\putbib[M:/Research/bib] % for use with bibunits
%\end{bibunit}

\end{document}